\documentclass[structabstract]{aa}
\usepackage{txfonts}

\usepackage[utf8]{inputenc}

\usepackage{natbib}
\bibpunct{(}{)}{;}{a}{}{,}

\usepackage[english]{babel}

\usepackage[section]{placeins}
\usepackage{multirow}

\usepackage{graphicx}
\usepackage{pst-pdf}
\usepackage{float}

\usepackage{array}
\usepackage{nicefrac}

\newcommand{\tfm}[1]{\tablefootmark{#1}}

\newcommand{\bigCellAlt}[2]{\parbox{#1}{\smallskip #2\smallskip}}

\newcommand{\inclGrII}[1]{\includegraphics{#1.pdf}}
\newcommand{\inclGrPNG}[1]{\includegraphics[width=\linewidth]{#1.png}}

\newcommand{\term}[1]{#1}
\newcommand{\termIrrel}[1]{#1}
\newcommand{\nth}[1]{$#1$th}
\newcommand{\chk}{$\times$} 

\newcommand{\para}[1]{\paragraph{#1}}

\newcommand{\eg}{e.~g.}
\newcommand{\ie}{i.~e.}
\newcommand{\viz}{viz.}
\newcommand{\wrt}{with respect to}

\newcommand{\lbl}{\label}
\newcommand{\refWordSec}{Section}
\newcommand{\refWordEq}{Eq.}
\newcommand{\refWordEqs}{Eqs.}
\newcommand{\refWordFig}{Fig.}
\newcommand{\refWordFigs}{Figs.}
\newcommand{\refWordTab}{Table}
\newcommand{\refWordTabs}{Tables}
\newcommand{\RefWordSec}{Section}
\newcommand{\RefWordEq}{Equation}

\newcommand{\RefWordFig}{Figure}

\newcommand{\RefWordTab}{Table}

\newcommand{\secref}[1]{\refWordSec~\ref{#1}}
\newcommand{\Secref}[1]{\RefWordSec~\ref{#1}}
\newcommand{\eqrf}[1]{\ref{#1}}
\newcommand{\eqref}[1]{\refWordEq~\eqrf{#1}}
\newcommand{\Eqref}[1]{\RefWordEq~\eqrf{#1}}
\newcommand{\eqsrefii}[2]{\refWordEqs\ \eqrf{#1} and \eqrf{#2}}

\newcommand{\eqsrefiii}[3]{\refWordEqs\ \eqrf{#1}, \eqrf{#2} and \eqrf{#3}}

\newcommand{\eqsrefrange}[2]{\refWordEqs\ \eqrf{#1} -- \eqrf{#2}}
\newcommand{\figref}[1]{\refWordFig~\ref{#1}}
\newcommand{\Figref}[1]{\RefWordFig~\ref{#1}}

\newcommand{\figsrefrange}[2]{\refWordFigs\ \ref{#1} -- \ref{#2}}

\newcommand{\tabref}[1]{\refWordTab~\ref{#1}}
\newcommand{\Tabref}[1]{\RefWordTab~\ref{#1}}
\newcommand{\tabsrefii}[2]{\refWordTabs\ \ref{#1} and \ref{#2}}

\newcommand{\citepeg}[1]{\citep[\eg][]{#1}}

\newcommand{\const}{\mathrm{const}}

\newcommand{\defby}{\equiv}
\newcommand{\vect}[1]{\vec{#1}}

\newcommand{\ddd}{\,\mathrm{d}}

\newcommand{\TMin}{{\mathrm{min}}}
\newcommand{\TMax}{{\mathrm{max}}}

\newcommand{\medSym}{\tilde}
\newcommand{\med}[1]{\medSym{#1}}

\newcommand{\mdSym}{\hat}
\newcommand{\margMdSym}{\check}
\newcommand{\md}[1]{\mdSym{#1}}
\newcommand{\margMd}[1]{\margMdSym{#1}}
\newcommand{\mnSym}{\bar}
\newcommand{\mn}[1]{\mnSym{#1}}

\newcommand{\intvlSym}{I}
\newcommand{\intvl}[1]{\intvlSym_{#1}}

\newcommand{\HPDI}{\intvl{\mathrm{HPD}}}
\newcommand{\HPDIArg}[1]{\intvl{\mathrm{HPD},\;#1}}

\newcommand{\EE}[2]{#1\cdot10^{#2}}
\newcommand{\EEi}[1]{10^{#1}}

\newcommand{\seti}[1]{\left\{#1\right\}}
\newcommand{\oneHalf}{\frac{1}{2}}
\newcommand{\tmDeriv}[1]{\dot{#1}}

\newcommand{\Real}{\mathbb R}
\newcommand{\Natural}{\mathbb N}

\newcommand{\winWidth}{\sigma_{\mathrm{ker}}}
\newcommand{\sampStdDev}{\sigma_{\mathrm{samp}}}
\newcommand{\sampStdDevPer}{\sigma'}
\newcommand{\interQuartRge}{r_{\mathrm{iq}}}
\newcommand{\kerEst}{\mathcal F}
\newcommand{\kr}{K}

\newcommand{\sgn}{\mathrm{sgn}}
\newcommand{\stdNormDistr}{\mathcal{N}(0,1)}

\newcommand{\corrCoeff}[2]{r_{#1,#2}}

\newcommand{\matrx}[1]{\tens{#1}}
\newcommand{\matrxi}[1]{
	\begin{array}{c}
		#1
	\end{array}
}
\newcommand{\pmatrxi}[1]{
	\left(\!\matrxi{#1}\!\right)
}
\newcommand{\matrxii}[1]{
	\begin{array}{cc}
		#1
	\end{array}
}
\newcommand{\pmatrxii}[1]{
	\left(\!\matrxii{#1}\!\right)
}
\newcommand{\matrxiii}[1]{
	\begin{array}{ccc}
		#1
	\end{array}
}
\newcommand{\pmatrxiii}[1]{
	\left(\!\matrxiii{#1}\!\right)
}

\newcommand{\transp}{^\mathrm T}
\newcommand{\inv}{^{-1}}
\newcommand{\niceSqrt}{^{\nicefrac{1}{2}}}

\newcommand{\rotmi}[1]{\matrx R(#1)}

\newcommand{\rotmzxz}{\matrx R_{zxz}}

\newcommand{\BASElong}{\textbf Bayesian \textbf astrometric and \textbf spectroscopic \textbf exoplanet detection and characterisation tool}

\newcommand{\I}{\mathcal K}
\newcommand{\mdl}{\mathcal M}
\newcommand{\mdlFct}{f}

\newcommand{\mdlFctAM}[2]{\vr(#1;#2)}
\newcommand{\mdlFctRV}[2]{\RV(#1;#2)}
\newcommand{\p}{\mathrm{p}}
\newcommand{\postr}{\mathcal P}
\newcommand{\pPri}[1]{\p(#1|\mdl,\I)}
\newcommand{\pLikh}[2]{\p(#1|#2,\mdl,\I)}
\newcommand{\pEv}[1]{\p(#1|\mdl,\I)}
\newcommand{\pPostr}[1]{\p(#1|\Data,\mdl,\I)}
\newcommand{\mapo}[2]{\postr_{#1}(#2)}
\newcommand{\jointmapo}[4]{\postr_{#1,#2}(#3,#4)}
\newcommand{\likhood}{\mathcal L}
\newcommand{\hy}{\mathcal H}
\newcommand{\probCont}{C}

\newcommand{\uniformPrior}[3]{\frac{\Theta(#1-#2)\,\Theta(#3-#1)}{#3-#2}} 
\newcommand{\jeffreysPrior}[3]{\frac{\Theta(#1-#2)\,\Theta(#3-#1)}{#1\ln\frac{#3}{#2}}} 
\newcommand{\modJeffreysPrior}[4]{\frac{\Theta(#1-#3)\,\Theta(#4-#1)}{(#1+#2)\ln\frac{#4+#2}{#2}}} 

\newcommand{\gravConst}{G}

\newcommand{\UNITd}{\mbox{d}}

\newcommand{\UNITmpsFlat}{\mathrm {ms}^{-1}}

\newcommand{\UNITkmpsFlat}{\mbox{kms}^{-1}}
\newcommand{\UNITmas}{\mbox{mas}}
\newcommand{\UNITmuas}{\mbox{$\mu$as}}
\newcommand{\UNITrad}{\mbox{rad}}
\newcommand{\UNITarcsecTxt}{\mbox{arcsec}}
\newcommand{\UNITAU}{\mbox{AU}}
\newcommand{\UNITkAU}{\mbox{kAU}}

\newcommand{\UNITdInvFlat}{\UNITd^{-1}}
\newcommand{\UNITyr}{\mbox{yr}}
\newcommand{\UNITpc}{\mbox{pc}}
\newcommand{\UNITMSun}{M_{\sun}}
\newcommand{\UNITMJup}{M_{\mathrm J}}
\newcommand{\SBII}{SB~II}
\newcommand{\sysS}{S}
\newcommand{\sys}[1]{$\sysS_#1$}

\newcommand{\vr}{\vect r}
\newcommand{\vv}{\vect v}

\newcommand{\errNomi}[1]{\epsilon_{0,#1}}

\newcommand{\errAddi}[1]{\epsilon_{+,#1}}

\newcommand{\errToti}[1]{\epsilon_{#1}}

\newcommand{\MCMClong}{Markov chain Monte Carlo}
\newcommand{\nSamp}{N}
\newcommand{\nSampBurn}{M}
\newcommand{\iSamp}{j}                
\newcommand{\samp}{\smp{\iSamp}}      
\newcommand{\smp}[1]{^{(#1)}}         

\newcommand{\PSR}{\hat{\mathcal R}^{\nicefrac{1}{2}}}

\newcommand{\PSRthr}{\mathcal R_\ast^{\nicefrac{1}{2}}}
\newcommand{\PSRthrSq}{\mathcal R_\ast}
\newcommand{\PSRSq}{\hat\mathcal R}
\newcommand{\PSRDOF}{d}
\newcommand{\PSRV}{\hat V}
\newcommand{\PSRW}{W}
\newcommand{\PSRB}{B}
\newcommand{\iniSamp}{\smp{0}}

\newcommand{\nxtSamp}{\smp{\iSamp+1}}

\newcommand{\accProb}[2]{\alpha(#1, #2)}
\newcommand{\tPar}{\gamma}
\newcommand{\nCh}{n}
\newcommand{\nPT}{m}
\newcommand{\iCh}{k}
\newcommand{\iPT}{l}
\newcommand{\indxPT}[1]{_{#1}}
\newcommand{\indexPT}{\indxPT{\iPT}}
\newcommand{\indxCh}[1]{_{(#1)}}
\newcommand{\indexCh}{\indxCh{\iCh}}
\newcommand{\nSwap}{n_{\mathrm{swap}}}

\newcommand{\RV}{v}
\newcommand{\param}{\theta}
\newcommand{\vPar}{\vect \param}

\newcommand{\vParRVCompoTxt}{$\PMe,\PMf,\PMchi,\PMomegaii,\PMV,\PMK_1$ and $\PMK_2$}

\newcommand{\vParAllCompo}{\PMe,\PMf,\PMchi,\PMomegaii,\PMi,\PMOmega,\PMpi,\PMV,\PMK_1,\PMK_2}

\newcommand{\vDQAllCompo}{\PMP,\PMT,\PMd,\PMrho,\PMKLiti{1},\PMKLiti{2},\PMmC{1},\PMmC{2},\PMaC{1},\PMaC{2},\PMaRel}

\newcommand{\EA}{E}
\newcommand{\MA}{M}

\newcommand{\knee}[1]{#1_{\mathrm k}}

\newcommand{\DELOmega}{\Delta\Omega}
\newcommand{\DELi}{\Delta i}
\newcommand{\DELomega}{\Delta\omega}

\newcommand{\DELangPos}{\Delta\AMAngSym}
\newcommand{\propMotAl}{\mu_{\alpha\cos\delta}}
\newcommand{\propMotDel}{\mu_\delta}
\newcommand{\propMotOrth}{\mu}

\newcommand{\vHPar}{\vect{\hat\param}}
\newcommand{\parSpace}{\Theta}
\newcommand{\nPars}{k}
\newcommand{\trueAnom}{\nu}

\newcommand{\PMV}{V}

\newcommand{\PMe}{e}
\newcommand{\PMf}{f}
\newcommand{\PMchi}{\chi}
\newcommand{\PMomegaS}{\omega_{\star}}
\newcommand{\PMomegaP}{\omega}

\newcommand{\PMomegaii}{\omega_2}

\newcommand{\PMK}{K}
\newcommand{\PMP}{P}
\newcommand{\PMT}{T}
\newcommand{\PMmP}{m_{\mathrm p}}
\newcommand{\PMmPi}[1]{m_{\mathrm p,#1}}
\newcommand{\PMmS}{m_\star}
\newcommand{\PMmC}[1]{m_{#1}}
\newcommand{\PMmmin}{\PMmPi{\mathrm{min}}}
\newcommand{\PMaS}{a_\star}
\newcommand{\PMaP}{a_{\mathrm p}}
\newcommand{\PMaC}[1]{a_{#1}}
\newcommand{\PMaApp}{a'}
\newcommand{\PMaRelApp}{a'_{\mathrm{rel}}}
\newcommand{\PMaRel}{a_{\mathrm{rel}}}
\newcommand{\PMaPMax}{a_{\mathrm{p,max}}}
\newcommand{\PMasini}{\PMaS\sin\PMi}
\newcommand{\PMKLit}{\PMK_{\mathrm{alt}}}
\newcommand{\PMKLiti}[1]{\PMK_{\mathrm{alt},#1}}
\newcommand{\PMKi}{K_1}
\newcommand{\PMKii}{K_2}
\newcommand{\PMi}{i}
\newcommand{\PMOmega}{\Omega}
\newcommand{\PMpi}{\varpi}
\newcommand{\PMd}{d}
\newcommand{\PMrho}{\rho}
\newcommand{\PMsgmP}{\sigma_+}
\newcommand{\PMtauP}{\tau_+}

\newcommand{\POSal}{\alpha}
\newcommand{\POSdel}{\delta}
\newcommand{\POSalCosDel}{\POSal\cos\POSdel}
\newcommand{\POSrho}{\rho}
\newcommand{\POStheta}{\theta}
\newcommand{\UNCERTELLa}{a}
\newcommand{\UNCERTELLb}{b}
\newcommand{\UNCERTELLphi}{\phi}
\newcommand{\RESIDphi}{\varphi}
\newcommand{\RESIDnorm}{\varrho}
\newcommand{\AMPosNum}[1]{\vr_{#1}}
\newcommand{\AMAngSym}{r}
\newcommand{\AMAngPosSym}{\vect\AMAngSym}
\newcommand{\AMAngPos}{\AMAngPosSym}
\newcommand{\AMAngPosNum}[1]{\AMAngPosSym_{#1}}

\newcommand{\tmSpan}{\Delta\tm}
\newcommand{\Data}{D}

\newcommand{\tm}{t}

\newcommand{\tmOne}{\tm_1}

\newcommand{\iAM}{_{\mathrm{AM}}}
\newcommand{\iRV}{_{\mathrm{RV}}}
\newcommand{\datum}{y}
\newcommand{\vdatum}{\vect y}
\newcommand{\measUnc}{\sigma}

\newcommand{\measUncNomi}[1]{\sigma_{0,#1}}
\newcommand{\measUncToti}[1]{\sigma_{#1}}
\newcommand{\dataCovM}{\matrx E}

\newcommand{\dataCovMNomi}[1]{\matrx E_{0,#1}}
\newcommand{\dataCovMToti}[1]{\matrx E_{#1}}
\newcommand{\dataCovMAdd}{\matrx E_+}
\newcommand{\cov}{\mathrm{cov}}
\newcommand{\var}{\mathrm{var}}

\newcommand{\clph}{\phi_{\mathrm{cl}}} 

\newcommand{\miz}{Mizar~A}
\newcommand{\mizb}{Mizar~B}
\newcommand{\mizAltNames}{\object{$\zeta^1$ Ursae Majoris}, \object{HD~116656}, \object{HR~5054}}
\newcommand{\runDesc}{pass}
\newcommand{\RunDesc}{Pass}
\newcommand{\run}[1]{\runDesc~#1}
\newcommand{\Run}[1]{\RunDesc~#1}

\newcommand{\vParTrue}{\vect \param_{\mathrm{true}}}

\begin{document}

%
%

	\title{Bayesian analysis of exoplanet and binary orbits\thanks{BASE, the computer program introduced in this article, can be downloaded at \texttt{http://www.mpia.de/homes/schulze/base.html}.}}
	\subtitle{Demonstrated using astrometric and radial-velocity data of \object{\miz}}
	\date{Received \ldots; accepted \ldots}
	\abstract
	{}
	{We introduce BASE (Bayesian astrometric and spectroscopic exoplanet detection and characterisation tool), a novel program for the combined or separate Bayesian analysis of astrometric and radial-velocity measurements of potential exoplanet hosts and binary stars. The capabilities of BASE are demonstrated using all publicly available data of the binary \miz.}
	{With the Bayesian approach to data analysis we can incorporate prior knowledge and draw extensive posterior inferences about model parameters and derived quantities. This was implemented in BASE by \MCMClong\ (MCMC) sampling, using a combination of the Metropolis-Hastings, hit-and-run, and parallel-tempering algorithms to explore the whole parameter space. Nonconvergence to the posterior was tested by means of the Gelman-Rubin statistic (potential scale reduction). The samples were used directly and transformed into marginal densities by means of kernel density estimation, a ``smooth'' alternative to histograms. We derived the relevant observable models from Newton's law of gravitation, showing that the motion of Earth and the target can be neglected.}
	{With our methods we can provide more detailed information about the parameters than a frequentist analysis does. Still, a comparison with the \miz\ literature shows that both approaches are compatible within the uncertainties.}
	{We show that the Bayesian approach to inference has been implemented successfully in BASE, a flexible tool for analysing astrometric and radial-velocity data.}
	\author{T. Schulze-Hartung\inst{\ref{mpia}} \and\ R. Launhardt\inst{\ref{mpia}} \and\ T. Henning\inst{\ref{mpia}}}
	\institute{Max-Planck-Institut für Astronomie, Königstuhl 17, D-69117 Heidelberg, Germany\\\email{schulze@mpia.de}}\label{mpia}
	\maketitle

%
	\section{Introduction}\lbl{sec_introduction}
%
	
		
		The search for extrasolar planets -- places where one day humankind might find other forms of life in the Universe -- has been a subject of scientific investigation since the nineteenth century, but only became successful in 1992 with the first confirmed discovery of an exoplanet orbiting the pulsar PSR~B1257+12 \citep{190}. Still, because it is situated in an environment hostile to life as we know it, this case has been of less relevance to the public than the first detection of a Sun-like planet-host star, 51 Pegasi \citep{191}. Since that time, more than 700 extrasolar planet candidates have been unveiled in more than 500 systems, more than 90 of which show signs of multiplicity \citep{192}.
		
		Closely related to the \term{detection} of extrasolar planets is the \term{characterisation} of their orbits. Both these tasks now profit from the existence of a variety of observational techniques, which we briefly sketch in the following. Comprehensive reviews can be found in \citet{434} and \citet[chapter 1]{433}.

		
		\term{Direct} observational methods refer to the \term{imaging} of exoplanets \citepeg{441}, which reflect the light of their host stars but also emit their own thermal radiation. To overcome the major obstacle of the high brightness contrast between planet and star, techniques such as \term{coronagraphy} \citep{483,441}, \term{angular differential imaging} \citep{480,487}, \term{spectral differential imaging} \citep{486,487}, and \term{polarimetric differential imaging} \citep{485,488} have been invented. Still, imaging has only revealed few detections and orbit determinations so far.

		
		The most productive methods in terms of the number of detected and characterised exoplanets are of an \term{indirect} nature, observing the effects of the planet on other objects or their radiation.


		Of these, \term{transit photometry} \citepeg{484,444} is noteworthy because it has helped uncover more than 200 exoplanet candidates, plus over 2,000 still unconfirmed candidates from the Kepler space mission \citep{641}: small decreases in the apparent visual brightness of a star during the \term{primary} or \term{secondary eclipse} point to the existence of a transiting companion. These data allow one to determine the planet's radius and orbital inclination and may also yield information on the planet's own radiation.

			
		Timing methods include measurements of \term{transit timing variations (TTV)} and \term{transit duration variations (TDV)} \citepeg{445,446} of either binaries or stars known to harbour a transiting planet. The method used in the first exoplanet detection \citep{190} is \term{pulsar timing}, which relies on slight anomalies in the exact timing of the radio emission of a pulsar and is sensitive to planets in the Earth-mass regime.


		\term{Microlensing} \citep{489,440}, which accounted for about 15 exoplanet candidates, uses the relativistic curvature of spacetime due to the masses of both a lens star and its potential companion, with the latter causing a change in the apparent magnification and thus the observed brightness of a background source.
		

		Perhaps the most well-known technique, and one of those on which this article is based, is known as \term{Doppler spectroscopy} or \term{radial-velocity} (RV) measurements \citepeg{191,490}. With more than 500 exoplanet candidates, it has been most successful in detecting new exoplanets and determining their orbits to date. From a set of high-resolution spectra of the target star, a time series of the line-of-sight velocity component of the star is deduced. These data allow one to determine the orbit in terms of its geometry and kinematics in the \term{orbital plane} as well as the minimum planet mass~$\PMmmin\approx\PMmP\sin\PMi$.
		To derive the actual planet mass~$\PMmP$, the inclination~$\PMi$ of the orbit plane \wrt\ the sky plane needs to be derived with a different method, \eg\ astrometry. The RV technique is distance-independent by principle, but signal-to-noise requirements do pose constraints on the maximum distance to a star. Stellar variability sometimes makes this approach difficult because it alters the line shapes and thus mimicks RV variations.
		The signal in stellar RVs caused by a planet in a circular orbit has a semi-amplitude of approximately
		\begin{equation}
			\PMK \approx \PMmP\sin\PMi\sqrt{\frac{G}{\PMmS \PMaRel}},\lbl{eq_RV_semiamp_general}
		\end{equation}
		where $\PMmP,\PMmS,\PMi,G,$ and $\PMaRel$ are the masses of planet and host star, the orbital inclination, Newton's gravitational constant, and the semi-major axis of the planet's orbit relative to the star, respectively. This approximation holds for $\PMmP\ll\PMmS$, which is true in most cases. It should be noted that the sensitivity of the RV method decreases towards less inclined (more face-on) orbits, which is an example for the selection effects inherent to any planet-detection method.
		
			
		Finally, \term{astrometry} \citep[AM; \eg][]{492,030,491} -- on which this work is also based -- is the oldest observational technique known in astronomy: a stellar position is measured with reference to a given point and direction on sky. Astrometry can thus be considered as complementary to Doppler spectroscopy, which measures the kinematics perpendicular to the sky plane. In contrast to the latter, AM allows one to determine the orientation of the orbital plane relative to the sky in terms of its inclination~$\PMi$ and the position angle~$\PMOmega$ of the \term{line of nodes} \wrt\ the meridian of the target.
		A planet in circular orbit around its host star displaces the latter on sky with an approximate angular semi-amplitude of
		\begin{equation}
			\alpha \approx \frac{\PMmP}{\PMmS}\frac{\PMaRel}{\PMd},\lbl{eq_AM_semiamp_general}
		\end{equation}
		where $\PMd$ is the distance between the star and the observer. Again, this approximation holds for $\PMmP\ll\PMmS$.
		
		Imaging astrometry, in its attempt to reach sufficient presicion, still faces problems due to various distortion effects. By contrast, \term{interferometric} astrometry has been used to determine the orbits of previously known exoplanets, mainly with the help of space-borne telescopes such as Hipparcos or the Hubble Space Telescope (HST), which presently still excel their Earth-bound competitors \citepeg{401}. However, instruments like PRIMA \citep{495,496,389} or GRAVITY \citep{424} at the ESO Very Large Telescope Interferometer are promising to advance ground-based AM even more in the near future.
		
		While planet-induced signals in AM and RVs are both approximately linear in planetary mass~$\PMmP$, they differ in their dependence on the orbital semi-major axis~$\PMaRel$ (\eqsrefii{eq_RV_semiamp_general}{eq_AM_semiamp_general}). Doppler spectroscopy is more sensitive to smaller orbits (or higher orbital frequencies, \eqref{eq_Kepler_III}), while AM favours larger orbital separations, \viz\ longer periods.

		
		In this article we introduce BASE, a \BASElong. Its goals are to fulfil two major tasks of exoplanet science, namely the \term{detection of exoplanets} and the \term{characterisation of their orbits}. BASE has been developed to provide for the first time the possibility of an integrated Bayesian analysis of stellar astrometric and Doppler-spectroscopic measurements \wrt\ their companions' signals,%
		\footnote{Although most methods described in this introduction apply to any kind of companion to a star, we refer here to companions as ``exoplanets'', irrespective of whether they are able to sustain hydrogen or deuterium burning.}
		correctly treating the measurement uncertainties and allowing one to explore the whole parameter space without the need for informative prior constraints.
		Still, users may readily incorporate prior knowledge, \eg\ from previous analyses with other tools, by means of priors on the model parameters.
		The tool automatically diagnoses convergence of its \MCMClong\ (MCMC) sampler to the posterior and regularly outputs status information.
		For orbit characterisation, BASE delivers important results such as the probability densities and correlations of model parameters and derived quantities.
		
		Because published high-precision AM observations of potential exoplanet host stars are still sparse, we used data of the well-known binary \miz\ to demonstrate the capabilities of BASE. It is also planned to gain astrophysical insights into exoplanet systems using BASE in the near future.

		
		This article is organised as follows. \Secref{sec_methodology} provides an overview of the most often-used methods of data analysis, including Bayes' theorem and MCMC as theoretical and implementational foundations of this work, as well as a derivation of the necessary observable models. BASE is described in \secref{sec_BASE}. In \secref{sec_target+data}, the target \miz\ and the data used in this article are discussed. \Secref{sec_results} presents and discusses our analysis of \miz. Conclusions are drawn in \secref{sec_conclus}.

%
	\section{Methods and models}\lbl{sec_methodology}
%

		
		Data analysis is a type of \term{inductive reasoning} in that it infers general rules from specific observational data \citepeg{075}. These general rules are described by \term{observable models}, simply called \term{models} in the following, which produce theoretical values of the observables as a function of \term{parameters}. The primary tasks of data analysis are listed in the following.
		
		\begin{enumerate}
			\item In \term{model selection}, the relative probabilities of a set of concurrent models~$\seti{\mdl_i}$, chosen a priori, are assessed. Specifically, \term{exoplanet detection} tries to decide the question of whether a certain star is accompanied by a planet or not, based on available data. Additionally, \term{model-assessment} techniques can be used to determine whether the most probable model describes the data accurately enough.
		
			\item \term{Parameter estimation} aims to determine the parameters~$\vPar$ of a chosen model. This is specifically referred to as \term{exoplanet characterisation} (or orbit determination) in the present context.
		
			\item The purpose of \term{uncertainty estimation} is to provide a measure of the parameters' uncertainties.
		\end{enumerate}

		
		Although model selection is equally important, in what follows we focus entirely on the second and third tasks, \viz\ parameter and uncertainty estimation. This is because for a known binary system, only one model is appropriate, viz. two bodies orbiting each other. Accordingly, BASE can only perform model selection when it analyses data from stars for which it is a-priori unknown whether a companion exists.

%
		\subsection{Likelihoods and frequentist inference}\lbl{sec_freq_inference}
%

			
			The well-established, conventional \term{frequentist} approach to inference is touched upon only briefly here. Its name stems from the fact that it defines probability as the relative frequency of an event. Measurements are regarded as values of \term{random variables} drawn from an underlying \term{population} that is characterised by \term{population parameters}, \eg\ mean and standard deviation in the case of a normal (Gaussian) population.
			%
			%
			%
			In the following, we derive the joint probability density of the values of AM and RV data, known as the \term{likelihood}~$\likhood$, which plays a central role in frequentist inference.
			
			When combining data of different types, one should generally be aware that potential systematic errors may differ between the data sets, \eg\ due to a calibration error in one instrument that renders the data inconsistent with each other. In this case, each data set analysed separately would imply a different result. In other instances, systematic errors in one set do not affect the other data: for example, any constant offset in radial velocity is absorbed into parameter~$\PMV$, which is irrelevant to the analysis of astrometric data. In the following derivation, we assumed that no systematic effects are present that led to inconsistent data.

			
			In the following, we assumed that the error~$\errToti{i}$ of any datum~$\datum_i$ is statistically independent of those of all other data and consists of two components, each distributed according to a (uni- or bivariate) normal distribution with zero mean:%
			\begin{itemize}
				\item a component~$\errNomi{i}$ corresponding to a nominal measurement error, whose distribution is characterised by the covariance matrix $\dataCovMNomi{i}$ or standard deviation~$\measUncNomi{i}$ given with the datum, for AM and RV data respectively, and
				\item a component~$\errAddi{i}$ representing \eg\ instrumental, atmospheric or stellar effects not modelled otherwise, whose distribution is characterised by scalar covariance matrix~$\dataCovMAdd = \mathrm{diag}(\PMtauP^2,\PMtauP^2)$ -- assuming no correlation between the noise in the two AM components -- or variance~$\PMsgmP^2$, respectively, where $\PMtauP$ and $\PMsgmP$ are free noise-model parameters.
			\end{itemize}
			The AM \term{data covariance matrix} $\dataCovMNomi{i}$ of datum~$i$, representing the uncertainty of and correlation between the two components measured, can be written using \termIrrel{singular-value decomposition} as
			\begin{equation}
				\dataCovMNomi{i} = \rotmi{-\UNCERTELLphi_i}\,\pmatrxii{
					a_i^2 & 0\\
					0 & b_i^2
				}\,\rotmi{\UNCERTELLphi_i},\lbl{eq_data_cov_matrix}
			\end{equation} 
			where $\rotmi{\cdot},a_i,b_i$ and $\UNCERTELLphi_i$ are the $2\times2$ \term{passive rotation matrix}, the nominal semi-major and -minor axes of the \term{uncertainty ellipse} and the position angle of its major axis, respectively.
			
			Using \termIrrel{characteristic functions}, it is readily shown that the sum $\errToti{i} = \errNomi{i}+\errAddi{i}$ of the two independent error components is again normally distributed, with zero mean and covariance matrix~$\dataCovMToti{i} = \dataCovMNomi{i}+\dataCovMAdd$ or standard deviation~$\measUncToti{i} = \left(\measUncNomi{i}^2+\PMsgmP^2\right)\niceSqrt$, respectively.

			The probability density%
			\footnote{Throughout, we use the term \term{probability density} wherever it refers to a continuous quantity, as opposed to \term{probability} for discrete quantities. \term{Probability distribution}, denoted by $\p(\cdot)$, is a generic term used for both cases.}
			of the values of $N\iAM$ two-dimensional AM data~$\seti{\vr_i}$ and $N\iRV$ RV data~$\seti{\RV_i}$, known as the \term{likelihood}~$\likhood$, is then given by
			\begin{equation}
				\likhood = \likhood\iAM\,\likhood\iRV,
			\end{equation}
			where $\likhood\iAM$ and $\likhood\iRV$ are the likelihoods pertaining to the individual data types,
			\begin{eqnarray}
				\likhood\iAM &=& \left((2\pi)^{N\iAM}\prod_{i=1}^{N\iAM}\sqrt{\det \dataCovM_i}\right)\inv \exp\left(-\oneHalf\chi^2\iAM\right),\\
				\likhood\iRV &=& \left((2\pi)^{\frac{N\iRV}{2}}\prod_{i=1}^{N\iRV}\measUncToti{i}\right)\inv \exp\left(-\oneHalf\chi^2\iRV\right).
			\end{eqnarray}
			Furthermore, the \term{sums of squares} $\chi^2\iAM$ and $\chi^2\iRV$ are defined by
			\begin{eqnarray}
				\chi^2\iAM &\defby& \sum_{i=1}^{N\iAM} (\vr_i-\mdlFctAM{\vPar}{\tm_i})\transp\,\dataCovM_i\inv\,(\vr_i-\mdlFctAM{\vPar}{\tm_i}),\\\lbl{eq_chi_sq_AM}
				\chi^2\iRV &\defby& \sum_{i=1}^{N\iRV} \left(\frac{\RV_i-\mdlFctRV{\vPar}{\tm_i}}{\measUncToti{i}}\right)^2,\lbl{eq_chi_sq_RV}
			\end{eqnarray}
			where $\vr_i,\RV_i$ are the \nth{i} AM and RV datum, $\mdlFctAM{\cdot}{\cdot},\mdlFctRV{\cdot}{\cdot}$ the AM and RV model functions and $\vPar$ is the vector of model parameters. The relevant models are derived in \secref{sec_obs_models}.

			\para{Parameter estimation.}
			Frequentist parameter estimation is generally equivalent to maximising $\likhood$ or minimising $\chi^2$ as functions of~$\vPar$. The resulting \term{best estimates} of the parameters~$\vHPar$ are therefore often called \term{maximum-likelihood} or \term{least-squares estimates}. For linear models, $\chi^2(\vPar)$ is a quadratic function and consequently $\vHPar$ can be found unambiguously by matrix inversion. In the more realistic cases of nonlinear models, however, $\chi^2$ may have many local minima, therefore care needs to be taken not to mistake a local minimum for the global one. Several methods exist to this end, including evaluation of $\chi^2(\vPar)$ on a finite grid, \term{simulated annealing} or \term{genetic algorithms} \citepeg{075}.

			\para{Uncertainty estimation.}
			Frequentist parameter uncertainties are usually quoted as \term{confidence intervals}. Procedures to derive these are designed such that when repeated many times based on different data, a certain fraction of the resulting intervals will contain the true parameters.
			Popular methods use \term{bootstrapping} \citep{524} or the \term{Fischer information matrix}, which is based on a local linearisation of the model \citepeg{096}. However, these methods suffer from specific caveats: the Fischer matrix is only appropriate for a quadratic-shaped $\chi^2$ in the vicinity of the minimum, and bootstrapping, which relies on modified data, may lead to severe misestimation of the parameter uncertainties, especially when these are large \citep{011}.
		
%
		\subsection{Bayesian inference}\lbl{sec_bayesian_statistics}
%
		
			
			\term{Bayesian} inference \citepeg{102}, which has gained popularity in various scientific disciplines during the past few decades, defines probability as the degree of belief in a certain \term{hypothesis}~$\hy$. While this is sometimes criticised as leading to subjective assignments of probabilities, Bayesian probabilities are not subjective if they are based on all relevant knowledge~$\I$, hence different people having the same knowledge will assign them the same value \citepeg{102}. Thus, Bayesian probabilities are conditional on the knowledge~$\I$, and this conditionality should be stated explicitly, as in the following equations.

%
			\subsubsection{Bayes' theorem}\lbl{sec_bayes_theorem}
%

				In the eighteenth century, Thomas Bayes laid the foundation of a new approach to inference with what is now known as Bayes' theorem \citep{464}. For the purpose of parameter and uncertainty estimation, the hypothesis~$\hy$ refers to the values of model parameters~$\vPar$, and Bayes' theorem can be expressed as
				\begin{equation}
					\pPostr{\vPar} = \frac{\pLikh{\Data}{\vPar}\cdot\pPri{\vPar}}{\pEv{\Data}},\lbl{eq_bayes_theorem_param_est}
				\end{equation}
				where $\p(\cdot)$ is a probability (density). Furthermore, $\Data \defby \seti{(\tm_i,\vdatum_i)}$ is the set of pairs of observational times%
				\footnote{In general, $\tm_i$ is the value of an \term{independent variable}, which may \eg\ be temporal or spatial. We assumed the measurement durations to be short in comparison with the characteristic time of orbital motion, given by the orbital period, and thus the observations to take place at points in time~$\tm_i$ which are known exactly.}
				and corresponding data values, and $\mdl$ denotes the particular model assumed. As mentioned above, all probabilities are also conditional on the knowledge~$\I$, including statements on the types of parameters and the parameter space~$\parSpace$ (which we assume to be a subset of $\Real^\nPars$ with $\nPars\in\Natural$) as well as the noise model.
				
				Using Bayes' theorem, the aim is to determine the \term{posterior} $\postr(\vPar) \defby \pPostr{\vPar}$, \ie\ the probability distribution of the parameters~$\vPar$ in light of the data~$\Data$, given the model~$\mdl$ and prior knowledge~$\I$. The other terms, located on the right-hand side of the theorem, are explained below.


				\begin{itemize}
					\item The term \term{prior} refers to the probability distribution $\pPri{\vPar}$ of the parameters~$\vPar$ given only the model and prior knowledge~$\I$; it characterises the knowledge about the parameters present before considering the data. For objective choices of priors, based on classes of parameters, see \secref{sec_priors}.

					\item The \term{likelihood} $\pLikh{\Data}{\vPar}$ is the probability distribution of the data values~$\Data$, given the times of observation, the model and the parameters. It is introduced in the context of frequentist inference in \secref{sec_freq_inference}.

					\item The \term{evidence} is the probability distribution of the data values~$\Data$, given the times of observation and the model but neglecting the parameter values,
					\begin{eqnarray}
						\pEv{\Data} &= &\int\p(\Data,\vPar|\mdl,\I)\ddd\vPar\lbl{eq_evidence}\\
						&=& \int\pPri{\vPar}\,\pLikh{\Data}{\vPar}\ddd\vPar.
					\end{eqnarray}
					It equals the integral of the product of prior and likelihood over the parameter space~$\parSpace$ and plays the role of a normalising constant, which is hard to calculate in practice, however.
				\end{itemize}
				
				It may be instructive here to note that the frequentist approach of maximising the likelihood~$\pLikh{\Data}{\vPar}$ is equivalent to maximising the posterior when assuming uniform priors~$\pPri{\vPar}$. This can be seen by inserting $\pPri{\vPar} = \const$ into \eqref{eq_bayes_theorem_param_est}, which leads to
				\begin{equation}
					\postr(\vPar) = \pPostr{\vPar} \propto \pLikh{\Data}{\vPar}.
				\end{equation}
				However, this maximum-likelihood approach ignores the fact that uniform priors are not always the most objective choice (see \secref{sec_priors}) and the posterior cannot be fully characterised just by the position of its maximum. Still, the latter can be used as a posterior summary in the Bayesian framework (\secref{sec_posterior_inference}).

%
			\subsubsection{Posterior inference}\lbl{sec_posterior_inference}
%
				
				\para{Sampling from the posterior.}
				To estimate the normalised posterior~$\postr(\vPar)$, \ie\ the probability distribution of the parameters~$\vPar$ in light of the data, $\nSamp$~samples of the parameter vector $\left\{\vPar\samp:\iSamp=1,\ldots,\nSamp\right\}$ are collected using the \term{\MCMClong} method \citep[\term{MCMC}; \eg][]{098} in the variant described by the \term{Metropolis-Hastings} algorithm \citep[MH;][]{550,548}, which performs a \termIrrel{random walk} through parameter space. The distribution of these samples -- excluding the first $\nSampBurn < \nSamp$ \term{burn-in} samples, which are still strongly correlated with the starting state~$\vPar\iniSamp$ -- converges to the posterior~$\postr(\cdot)$ in the limit of many samples if the chain obeys certain \term{regularity conditions} \citepeg{150}. Methods for setting the starting state~$\vPar\iniSamp$ and detecting convergence are described in \secref{sec_comput_techniques}.
				
				Starting from the current chain link~$\vPar\nxtSamp \in \parSpace$, the following steps lead to the next link, according to the MH algorithm and the \term{hit-and-run sampler} \citep[step~\ref{it_sample_hit+run};][]{599,600}:
				\begin{enumerate}
					\item Set up a candidate~$\vect C$:\lbl{it_sample_hit+run}
					\begin{enumerate}
						\item sample a direction, \viz\ a random unit vector~$\vect d\in\Real^\nPars$ from an isotropic density over the $\nPars$-dimensional unit sphere;
						\item sample a (signed) distance~$r$ from a uniform distribution over the interval~$\{r' \in \Real: \vPar\samp+r'\vect d \in \parSpace\}$;
						\item set candidate $\vect C \defby \vPar\samp+r\vect d$;
					\end{enumerate}
					\item calculate the acceptance probability\lbl{it_sample_accept}
					\begin{equation}
						\accProb{\vPar\samp}{\vect C} \defby \min\left(1,\frac{\postr(\vect C)}{\postr(\vPar\samp)}\right);
					\end{equation} 
					\item draw a random number~$\beta$ from a uniform distribution over the interval~$[0,1]$;
					\item if $\beta \le \alpha$, accept the candidate, \ie\ set the next link $\vPar\nxtSamp \defby \vect C$; otherwise, $\vPar\nxtSamp \defby \vPar\samp$.
				\end{enumerate}
				The hit-and-run sampler, compared to alternatives like the \termIrrel{Gibbs sampler} \citep{552}, favours exploring of the whole parameter space~$\parSpace$ without becoming ``trapped'' in the vicinity of a local posterior maximum \citep{098}.
				
				Because only the ratio of two posterior values is used (step~\ref{it_sample_accept}), the normalising evidence~$\pEv{\Data}$ -- a constant that is difficult to determine, as mentioned above -- is irrelevant in the MH algorithm.

				\para{Marginalisation and density estimation.}
				Obviously, the posterior mode alone reveals only one particular aspect of the posterior. However, as a density over $\nPars > 2$ dimensions, the posterior cannot be displayed unambiguously in a figure.
				
				To obtain a plottable summary of the posterior, a set of \term{marginal posteriors}~$\mapo{i}{\cdot}$, \ie\ probability densities over each of the parameters~$\param_i$, and \term{joint marginal posteriors}~$\jointmapo{i}{j}{\cdot}{\cdot}$ over two parameters, can be estimated.
				Theoretically, these densities are derived from the posterior density by \term{marginalisation}, \viz\ integration over all other parameters,
				\begin{eqnarray}
					\mapo{i}{\param_i} &\defby& \pPostr{\param_i} = \int\postr(\vPar)\ddd\vPar_{\backslash i}\lbl{eq_def_mapo}\\
					\jointmapo{i}{j}{\param_i}{\param_j} &\defby& \pPostr{\param_i,\param_j} = \int\postr(\vPar)\ddd\vPar_{\backslash i,j},\lbl{eq_def_JMP}
				\end{eqnarray}
				where $\ddd\vPar_{\backslash i} \defby \prod_{k\neq i}\ddd\param_k$ and $\ddd\vPar_{\backslash i,j} \defby \prod_{k\neq i,j}\ddd\param_k$.
				Practically, marginal posteriors are estimated by only considering component~$i$ of the collected samples~$\seti{\vPar\samp}$ and performing a \term{density estimation} based on them. Joint marginal posteriors are derived analogously, based on components~$i$ and $j$.

				Several \term{density estimators} exist for deriving a density from a set of samples. One of them -- the oldest and probably most popular type, known as the \termIrrel{histogram} -- has several drawbacks: its shape depends on the choice of \termIrrel{origin} and \termIrrel{bin width}, and when used with two-dimensional data, a contour diagram cannot easily be derived from it. Generalising the histogram to \term{kernel density estimation} over one or two dimensions, the samples can be represented more accurately and unequivocally \citep{404}.
				
				Below, we refer only to the simpler one-dimensional case. There, the \termIrrel{kernel estimator} can be written as
				\begin{equation}
					\kerEst(x) \defby \frac{1}{\nSamp\winWidth}\sum_{\iSamp=1}^{\nSamp}\kr\left(\frac{x-X\samp}{\winWidth}\right),
				\end{equation}
				where $x$ is a scalar variable, $\kr(\cdot)$ the \term{kernel}, $\winWidth$ the \term{window width} and $X\samp$ are the underlying samples.
				As detailed by \citet{404}, the \termIrrel{efficiency} of various kernels in terms of the achievable \termIrrel{mean integrated square error} is very similar, and therefore the choice of kernel can be based on other requirements. Since no differentiability is required for the estimated densities and computational effort plays an important practical role, a \termIrrel{triangular} kernel,
				\begin{equation}
					\kr_{\mathrm{tri}}(x) \defby \max\left(1-|x|,0\right),
				\end{equation}
				was selected for estimating the marginal posteriors.
				The window width is chosen following the recommendations of \citet{404},
				\begin{equation}
					\winWidth \defby 2.189\cdot\min\left(\sampStdDev,\frac{\interQuartRge}{1.34}\right)\,(\nSamp-\nSampBurn)^{\frac{1}{5}},\lbl{eq_def_KDE_win_width}
				\end{equation}
				where $\sampStdDev,\interQuartRge,\nSamp$ and $\nSampBurn$ are the \termIrrel{sample standard deviation}, the \termIrrel{interquartile range} of the samples, number of samples and burn-in length, respectively.

				\para{Periodogram mode.}
				By default, the window width for marginal posteriors is based on the MCMC sample standard deviation~$\sampStdDev$ (\eqref{eq_def_KDE_win_width}). If there are multiple maxima, however, this can lead to artificially broad peaks, which can be particularly problematic for the orbital frequency~$\PMf$ (see \secref{sec_obs_models}), which plays an important role in distinguishing different solutions in orbit-related parameter estimation. Therefore, BASE includes a \term{periodogram mode}, in which the window width of the marginal posterior of $\PMf$ (a Bayesian analogon to the frequentist \term{periodogram}) is reduced according the following procedure.
				\begin{enumerate}
					\item Initially assume a default window width as given by \eqref{eq_def_KDE_win_width};
					\item\lbl{it_periodogram_find_min_max} estimate the marginal posterior of $\PMf$ and find its local maximum $\PMf_\TMax$ nearest to posterior mode~$\md\PMf$, as well as the local minimum $\PMf_\TMin$ nearest to $\PMf_\TMax$;
					\item\lbl{it_periodogram_calc_peak_stdev} calculate the marginal-posterior standard deviation $\sampStdDevPer$ over the half-peak between $\PMf_\TMax$ and $\PMf_\TMin$;
					\item\lbl{it_periodogram_recalc_win_width} re-calculate the window width using \eqref{eq_def_KDE_win_width}, with $\sqrt{2}\sampStdDevPer$ replacing $\sampStdDev$;
					\item repeat step~\ref{it_periodogram_find_min_max}, but only consider local minima with ordinates~$\pPostr{\PMf_\TMin} \leq 0.5\cdot\pPostr{\PMf_\TMax}$ in order not to be misled by weak marginal-posterior fluctuations;
					\item repeat steps \ref{it_periodogram_calc_peak_stdev} and \ref{it_periodogram_recalc_win_width}.
				\end{enumerate}

				\para{Parameter estimation.}
				To obtain a single most probable estimate of the parameters, the posterior density~$\postr(\cdot)$ can be summarised by the \term{posterior mode}~$\md\vPar\in\parSpace$, \ie\ the point where the posterior assumes its maximum value,
				\begin{equation}
					\md\vPar \defby {\arg\max}_{\vPar}\;\postr(\vPar).
				\end{equation} 
				This point, also known as the \term{maximum a-posteriori (MAP)} parameter estimate, can be approximated by the MCMC sample with highest posterior density, based on the values of $\postr(\vPar\samp)$ already calculated during sampling. This approximation neglects the finite spacing between samples.
				
				Alternatively, the following scalar summaries can be inferred from the samples or, for the marginal mode, from the marginal posteriors~$\mapo{i}{\param}$:
				\begin{itemize}
					\item \term{mean} or \term{expectation}~$\mn\param$,
					\begin{equation}
						\mn\param \defby \int_{-\infty}^{\infty} \param\,\postr_i(\param)\ddd\param,
					\end{equation}

					\item \term{median}~$\med\param$,
					\begin{equation}
						\int_{-\infty}^{\med\param} \postr_i(\param)\ddd\param \defby 0.5,
					\end{equation}
					
					\item \term{marginal mode} $\margMd\param$,
					\begin{equation}
						\margMd\param \defby {\arg\max}_\param\;\postr_i(\param).
					\end{equation}
				\end{itemize}

				\para{Uncertainty estimation.}
				
				For uncertainty estimation, \term{highest posterior-density intervals (HPDIs)} can be derived from the posterior samples. For any given $\probCont\in\Real$ with $0<\probCont<1$, a HPDI~$\HPDI \defby [a,b]$ is defined as the smallest interval over which the posterior contains a probability~$\probCont$,
				\begin{equation}
					\int_a^b\postr_i(\param)\ddd\param = \probCont, \quad\mathrm{s.t.}\quad b-a=\min.
				\end{equation} 
				BASE automatically calculates HPDIs of probability contents 50\%, 68.27\%, 95\%, 95.45\%, 99\%, and 99.73\%; others may be added on user request.
				
				In contrast to frequentist confidence intervals, HPDIs are generally not symmetric, meaning that their midpoint does not correspond to the best estimate. This is because the marginal posteriors may be asymmetric, including any amount of skew.
				
				It should also be noted that HPDIs are not useful with multimodal posteriors because several modes cannot be meaningfully summarised by one interval per dimension, nor by a single best estimate.
				
				To quantify linear dependencies between parameters, the a-posteriori Pearson correlation coefficient,
				\begin{equation}
					\corrCoeff{\param_1}{\param_2} \defby \frac{\cov(\param_1,\param_2)}{\sqrt{\var(\param_1)\,\var(\param_2)}} = \corrCoeff{\param_2}{\param_1},
				\end{equation}
				can be inferred from the samples. There may also be nonlinear correlations between parameters that are not described by the correlation coefficients. One should also be aware that for strong linear or non-linear relationships between parameters, uncertainties of single parameters as characterised by HPDIs may not be meaningful.
				
				We stress that the (joint) marginal posteriors can -- and should -- always be referred to, especially when best estimates and/or HPDIs do not adequately characterise the posterior. The availability of these more informative densities is one of the advantages of a Bayesian approach with posterior sampling.

%
		\subsection{Observable models}\lbl{sec_obs_models}
%
			
			Independent of the chosen approach to inference -- frequentist or Bayesian --, theoretical values of the observables need to be calculated and compared to the data by means of the likelihood. To this end, an observable model is set up for each relevant type of data, \ie\  a function $\mdlFct(\vPar; \tm)$ of the model parameters~$\vPar$ and time~$\tm$. An overview of all model parameters used in this work is given in \Tabref{tab_pars}, while \tabref{tab_DQs} lists quantities that can be derived from them.
			
			In this section, we only sketch the derivation of the observable models, beginning with a single-planet system. For an in-depth treatment of celestial mechanics, the interested reader is referred \eg\ to \citet{673}.

			
%
			\subsubsection{Stellar motion in the orbital plane}\lbl{sec_motion_in_orbital_plane}
%

				Newton's Law of Gravity governs the motion of a non-relativistic two-body system of star and planet, whose \term{centre of mass (CM)} rests in some \term{inertial} reference frame. A solution to it is given by both the star and the planet moving in elliptical \term{Keplerian orbits} with a fixed common \term{orbital plane} and each with one \term{focus} coinciding with the CM.
				
				To describe the stellar position, whose variation is observable with astrometry and Doppler spectroscopy, we set up a coordinate system~\sys{1} whose origin is identical to the CM, $z$-axis perpendicular to the orbital plane and the vector from the CM to the \term{periapsis} orientated in positive $x$-direction. In \sys{1}, the stellar barycentric position is given by
				\begin{equation}\lbl{eq_pos_star_S_one}
					\vr_1 = \PMaS
								\pmatrxi{
									\cos\EA-\PMe\\
									\sqrt{1-\PMe^2}\sin\EA\\
									0
								}
							= \vr_1(\EA;\PMaS,\PMe),
				\end{equation}
				where $\PMaS$ is the \term{semi-major axis}, $\PMe$ the \term{eccentricity} and $\EA$ the \term{eccentric anomaly}.
				
				The time-dependent eccentric anomaly is determined implicitly by \term{Kepler's equation},
				\begin{equation}
					\EA-\PMe\sin\EA = 2\pi(\PMchi+\PMf(\tm-\tmOne)) = \MA(\tm),\\\lbl{eq_Keplers_eq_chi}
				\end{equation}
				where $\PMf=\PMP\inv$ is the \term{orbital frequency}, $\PMP$ the \term{orbital period}, $\tmOne$ the time of first measurement and $\MA(\cdot)$ the \term{mean anomaly}, which varies uniformly over the course of an orbit. Furthermore, following \citet{007}, we use
				\begin{equation}
					\PMchi \defby \frac{\MA(\tmOne)}{2\pi} = \PMf(\tmOne-\PMT),\lbl{eq_def_chi}
				\end{equation}
				with $\PMT$ standing for the last time the periapsis was passed prior to $\tmOne$ (\term{time of periapsis}).
				Kepler's equation is \termIrrel{transcendental} and needs to be solved numerically to obtain $\EA$ for every relevant combination of $\PMe$ and $\MA$.
				
				BASE performs a one-time pre-calculation of $\EA$ over an $(\PMe,\MA)$-grid, which, because of the monotonicity of $\EA$ as an (implicit) function of $\PMe$ and $\MA$, allows one to reduce the effort of numerically solving \eqref{eq_Keplers_eq_chi} by providing lower and upper bounds on $\EA$.
				
				By reference to \eqsrefii{eq_pos_star_S_one}{eq_Keplers_eq_chi}, it is readily shown that the stellar coordinates are \termIrrel{periodic functions} of $\PMchi$ with period 1. We therefore call $\PMchi$ a \term{cyclic} parameter and treat it as lying in the range $[0,1)$.

	%
				\subsubsection{Transformation into the reference system}\lbl{sec_trafo_into_ref_sys}
	%

				\begin{figure}
					\centering
					\inclGrII{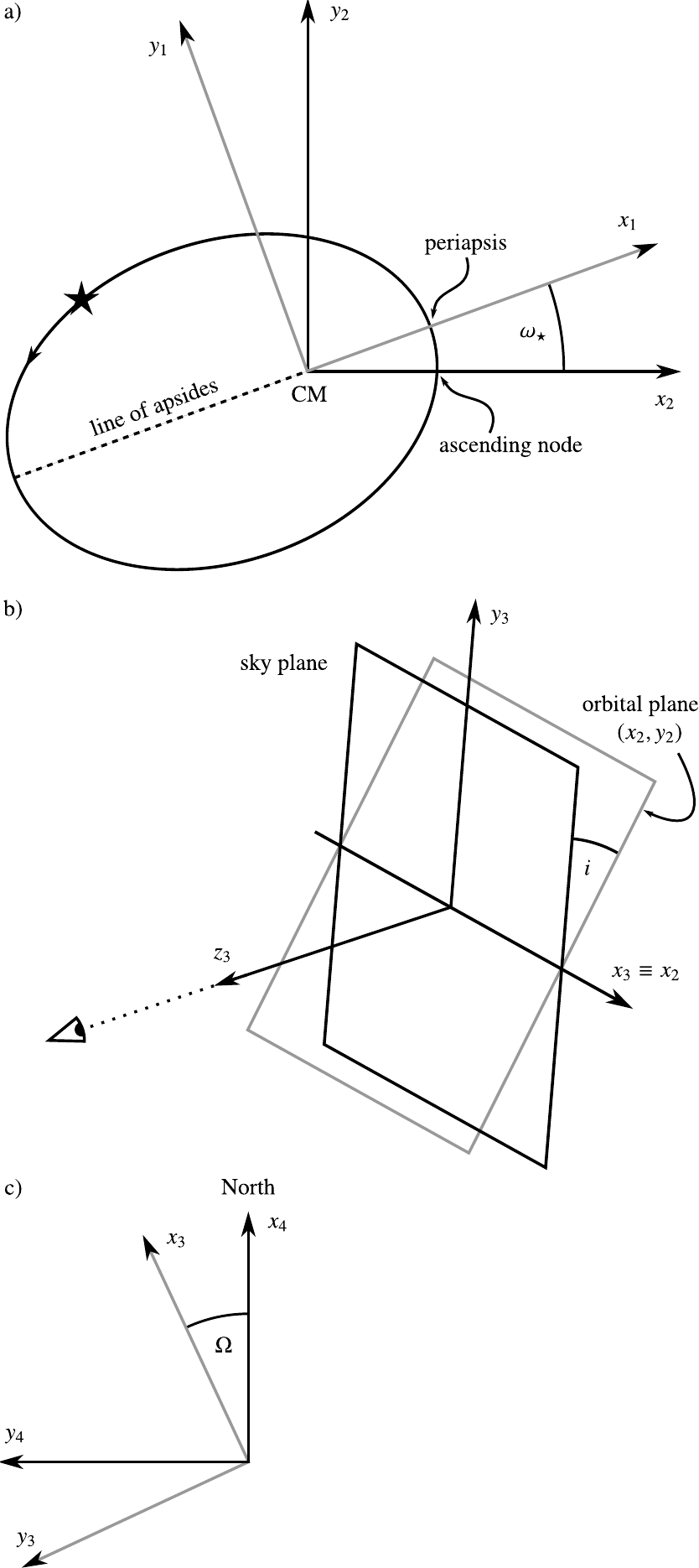}
					\caption{Definition of the angles $\PMomegaS,\PMi,\PMOmega$. a) From \sys{1} to \sys{2}, the star and its sense of rotation about the CM are indicated; the dotted line marks the major axis of the orbital ellipse. b) From \sys{2} to \sys{3}, the observer and line of sight are indicated. c) From \sys{3} to \sys{4}, the positive $x_4$-axis points northward along the meridian of the CM.}\lbl{fig_def_orientation_angles}
				\end{figure}

				To derive the stellar barycentric position as seen from the perspective of an observer, we transform \sys{1} into a new coordinate system~\sys{4} by three successive rotations. These are described by three \termIrrel{Euler angles}, termed in our case \term{argument of the periapsis}~$\PMomegaS$, \term{inclination}~$\PMi$ and \term{position angle of the ascending node}~$\PMOmega$, and are carried out as follows (\figref{fig_def_orientation_angles}):
				\begin{enumerate}
					\item Rotate \sys{1} about its $z_1$-axis by $(-\PMomegaS)$ such that the \term{ascending node}%
					\footnote{The \term{ascending node} is the point of intersection of the orbit and the sky plane where the moving object passes away from the observer. In step~2, a positive rotation angle is used to ensure that the node is indeed ascending
					(\figref{fig_def_orientation_angles}~b).}
					of the stellar orbit lies on the positive $x_2$-axis.
					
					\item Rotate \sys{2} about its $x_2$-axis by $(+\PMi)$ such that the new $z_3$-axis passes through the observer.
					
					\item Rotate \sys{3} about its $z_3$-axis by $(-\PMOmega)$ such that the new $x_4$-axis is parallel to the \term{meridian} of the CM and points in a northern direction.
				\end{enumerate}
				Thus, the stellar barycentric position has new coordinates
				\begin{equation}
					\vr_4 = \rotmzxz\,\vr_1,\lbl{eq_pos_star_S_four}
				\end{equation}
				with matrix
				\begin{equation}\lbl{eq_rot_matrix_comp}
					\rotmzxz \defby
					\pmatrxiii{
						A & F & J\\
						B & G & K\\
						C & H & L
					}
				\end{equation}
				defining the rotations; its components are
				\begin{eqnarray}
					A &=& \cos\PMOmega\cos\PMomegaS-\sin\PMOmega\cos\PMi\sin\PMomegaS\lbl{eq_rot_matrix_A}\\
					B &=& \sin\PMOmega\cos\PMomegaS+\cos\PMOmega\cos\PMi\sin\PMomegaS\\
					F &=& -\cos\PMOmega\sin\PMomegaS-\sin\PMOmega\cos\PMi\cos\PMomegaS\\
					G &=& -\sin\PMOmega\sin\PMomegaS+\cos\PMOmega\cos\PMi\cos\PMomegaS\lbl{eq_rot_matrix_G}\\
					J &=& -\sin\PMOmega\sin\PMi\\
					K &=& \cos\Omega\sin\PMi\lbl{eq_rot_matrix_K}\\
					C &=& -\sin\PMi\sin\PMomegaS\\
					H &=& -\sin\PMi\cos\PMomegaS\\
					L &=& \cos\PMi.
				\end{eqnarray}
				$A,B,F,$ and $G$ are known as the \term{Thiele-Innes constants}, first introduced by \citet{500}.

				
				By taking the time derivative of \eqref{eq_pos_star_S_four}, we obtain the stellar velocity in \sys{4},
				\begin{equation}
					\vv_4 = \rotmzxz\frac{\ddd\vr_1}{\ddd\EA}\tmDeriv{\EA} = \frac{2\pi\PMf\PMaS}{1-\PMe\cos\EA}\rotmzxz\pmatrxi{
						-\sin\EA\\
						\sqrt{1-e^2}\cos\EA\\
						0
					}.\lbl{eq_RV_S_four}
				\end{equation}

%
			\subsubsection{Relation to the planetary orbit}\lbl{sec_pl_orbit}
%

				By reference to the above results, the observables of AM and RV are easily derived (\secref{sec_observables}). Based on the following simple relation, they can be parameterised by quantities pertaining to the planetary instead of the stellar orbit.
				
				According to the definition of the CM, the line connecting star and planet contains the CM and the ratio of their respective distances from the CM equals the inverse mass ratio,
				\begin{equation}
					\overrightarrow{\mathrm{CS}} = -\frac{\PMmP}{\PMmS}\,\overrightarrow{\mathrm{CP}},\lbl{eq_def_CM}
				\end{equation}
				where C, S and P stand for CM, star and planet, respectively.
				This implies a simple relation between the orbits of the star and the planet as follows. The two bodies orbit the CM with a common orbital frequency~$\PMf$ and time of periapsis~$\PMT$. With regard to the corresponding periapsis, they always have the same eccentric anomaly~$\EA$. Their orbital shapes, \viz\ eccentricities~$\PMe$, are identical as well.
				
				Furthermore, the two bodies share the same sense of orbital revolution, hence those nodes of both orbits which lie on the positive $x_2$-axis are ascending. Consequently, the only Euler angle that differs between stellar and planetary orbit is the argument of periapsis, which differs by $\pi$ because star and planet are in opposite directions from the CM.

%
			\subsubsection{Observables}\lbl{sec_observables}
%

				To express the stellar barycentric position as a two-dimensional angular position, we performed a final transformation of \sys{4} into a spherical coordinate system \sys{5} with radial, elevation and azimuthal coordinates $(r,\POSdel,\POSal)$. Its origin is identical with the observer, its reference plane coincides with the $(y_4,z_4)$-plane and its fixed direction is $-z_4$. In \sys{5}, the radial coordinate of the CM equals a distance~$\PMd = 1\UNITAU\,\PMpi\inv$, with $\PMpi$ being the \term{parallax}.
				Stellar coordinates $\vr_4$ therefore correspond to a two-dimensional angular position
				\begin{equation}
					\AMAngPosNum{5} \defby \pmatrxi{
						\POSdel\\
						\POSal
					} = \frac{1}{\PMd}\,\pmatrxi{
						x_4\\
						y_4
					}
				\end{equation}
				in \sys{5}, where $\POSdel$ and $\POSal$ are called the \term{declination} and \term{right ascension}, respectively.
				Using \eqsrefiii{eq_pos_star_S_one}{eq_pos_star_S_four}{eq_rot_matrix_comp}, the model function for the angular position of the star \wrt\ the CM becomes
				\begin{equation}\lbl{eq_pos_star_S_five}
					\mdlFctAM{\vPar}{\tm} \defby \AMPosNum{5}
					= \PMaApp\left((\cos\EA-\PMe)\,
					\pmatrxi{
						A\\
						B
					}+\sqrt{1-\PMe^2}\sin\EA\,
					\pmatrxi{
						F\\
						G
					}\right),
				\end{equation}
				with $\PMaApp \defby \PMpi\PMaS(1\UNITAU)\inv$. In practice, complications arise because the stellar position is often measured relative to a physically unattached reference star, whose distance and motion differ, and not relative to the unobservable CM. By contrast, for a visual binary, the companion can be used as a reference, yielding the simple model described below.

				Other astrometric effects may be caused by the accelerated motion of Earth-bound observers around the \term{solar system barycentre (SSB)}. This is discussed for the case of the binary \miz\ in \secref{sec_motion_of_observer}.
			
				In contrast to astrometry, RV data are usually automatically transformed into an inertial frame resting \wrt\ the SSB \citepeg{511}, which allows the Earth's motion to be neglected in this model and treats the observer's rest frame as being inertial. The model function for the stellar radial velocity measured by an observer is thus given by $(\vv_5)_r$, with
				\begin{equation}
					(\vv_5)_r-\PMV = -(\vv_4)_z = \PMK\,\frac{\sin\EA\sin\PMomegaP-\sqrt{1-e^2}\cos\EA\cos\PMomegaP}{1-\PMe\cos\EA},\lbl{eq_RV_S_five}
				\end{equation}
				where $\PMV$ is the RV offset, consisting of the radial velocity of the CM plus an offset due to the specific calibration of the instrument -- which therefore differ, in general, between RV data sets -- and $\PMK$ is the \term{RV semi-amplitude}, which can be expressed as
				\begin{equation}
					K = 2\pi\PMf\PMaS\sin\PMi = \sqrt[3]{2\pi\gravConst\PMf}\,\PMmP(\PMmP+\PMmS)^{-\frac{2}{3}}\sin\PMi.\lbl{eq_RV_amplitude_II}
				\end{equation} 
				The last equality holds because of \term{Kepler's third law},
				\begin{equation}
					(\PMaP+\PMaS)^3\PMf^2 = \frac{\gravConst}{4\pi^2}\,(\PMmP+\PMmS)\lbl{eq_Kepler_III}
				\end{equation} 
				and the definition of the CM (\eqref{eq_def_CM}).
				Owing to \eqref{eq_RV_amplitude_II}, only one of $\PMaS,\PMK$ needs to be employed in the AM and RV models; for BASE, $\PMK$ was adopted.
				
				As an aside, in the literature \citepeg{007} often a different version of \eqref{eq_RV_S_five} is employed that involves $\trueAnom$ instead of $\EA$ and the alternative definition $\PMKLit \defby \frac{\PMK}{\sqrt{1-\PMe^2}}$ (see \tabref{tab_DQs}).
				
				\para{Binary system.}
				If the \term{primary} and \term{secondary} binary components assume the roles of star and planet, respectively, the above reasoning also yields the observables of a binary system.
				
				For visual binaries, AM measurements often refer to the position of the secondary \wrt\ the primary. From the definition of the CM, it follows that the orbit of the secondary \wrt\ the primary is identical with its barycentric orbit but scaled by a factor $(\PMmC{1}+\PMmC{2})\PMmC{1}\inv$, or with the semi-major axis equaling the sum of the two components' barycentric semi-major axes,
				\begin{equation}
					\PMaRel = \PMaC{1}+\PMaC{2}.\lbl{eq_rel_semimaj_ax_binary}
				\end{equation}
				Thus, the AM model of \eqref{eq_pos_star_S_five} can be used for a binary with $\PMaRel$ replacing $\PMaS$, $\PMomegaii+\pi$ replacing $\PMomegaS$ and
				\begin{equation}
					\PMaRelApp \defby \frac{\PMpi\PMaRel}{1\UNITAU}.\lbl{eq_def_a_rel_app}
				\end{equation}
				\Eqref{eq_RV_S_five} yields the RV of component~$i$ if $\PMK$ is replaced by $(-1)^{i+1} \PMK_i$ and $\PMomegaP$ by $\PMomegaii$.
				If AM and RV data are combined, BASE uses $\PMK_{1,2}$ instead of $\PMaC{1,2}$ and calculates $\PMaRel$ using the equivalent of \eqref{eq_RV_amplitude_II} in combination with \eqref{eq_rel_semimaj_ax_binary}.

%
			\subsubsection{Effects of the motion of observer and CM}\lbl{sec_motion_of_observer}
%

				If we consider the observer, whose position relative to the CM defines the orientation of \sys{{2\ldots5}}, to be located on Earth and therefore to be subject to Earth's accelerated motion around the SSB,
				the angles $\PMOmega,\PMi,\PMomegaP$ as defined above are not constant but rather functions of time; an additional source of their variation is the (assumedly linear) \term{proper motion} of the CM. Consequently, these angles are not appropriate constant model parameters even within a timespan of a few months. Furthermore, the systems \sys{{2\ldots5}} defined above are not strictly inertial, which renders the simple coordinate transformations invalid.
				However, we argue below that the variation in these angles is so weak that these effects are indeed negligible in the context of this work.
				
				Because AM determines instantaneous positions, it is directly influenced by changes in the positions of the observer and the CM. In the following, we assess for \miz\ the greatest possible magnitude of a change in the AM position $\DELangPos \defby |\Delta\AMAngPos|$ due to angular changes $\DELOmega,\DELi,\DELomega > 0$, which are in turn caused by the varying relative position of observer and CM.
				Finally, we compare $\DELangPos$ to the AM measurement uncertainties.
				
				An upper limit on each of the angular changes $\DELOmega,\DELi,\DELomega$ can be derived from the proper motion of the CM and the \term{annual parallax} from the Earth's motion, as
				\begin{equation}
					|\DELOmega|,|\DELi|,|\DELomega| \le 2\PMpi+\propMotOrth\,\tmSpan,\lbl{eq_max_ang_chg}
				\end{equation}
				where the first term corresponds to the annual parallax and $\propMotOrth$ is the magnitude of the CM's proper motion. For the second term, we have $\propMotOrth\tmSpan \approx \sqrt{\propMotAl^2+\propMotDel^2}\,\tmSpan$.
				
				Using $\sin(x+\xi) \approx \sin x + \xi \cos x$,
				for $\xi \ll 1$,
				along with \eqsrefrange{eq_rot_matrix_A}{eq_rot_matrix_K} and \eqrf{eq_max_ang_chg} as well as the final results for $\PMOmega,\PMi,\PMomegaP$ and $\PMpi$ (using the values~$\md\param$ from \tabref{tab_res_pars}) and the proper motion from \tabref{tab_miz_basic_props} yields the following maximum changes of the Thiele-Innes constants:
				\begin{eqnarray}
					|\Delta A| &\lessapprox& \EE{8.30}{-6}\\
					|\Delta B| &\lessapprox& \EE{8.00}{-6}\\
					|\Delta F| &\lessapprox& \EE{7.99}{-6}\\
					|\Delta G| &\lessapprox& \EE{4.34}{-6}.
				\end{eqnarray}
				
				Hence, with $\DELangPos \approx \sqrt{|\Delta\POSdel|^2+|\Delta(\POSalCosDel)|^2}$,
				\begin{eqnarray}
					|\Delta\POSdel| &\le& \PMaRelApp\,(2|\Delta A|+|\Delta F|)\\
					|\Delta(\POSalCosDel)| &\le& \PMaRelApp\,(2|\Delta B|+|\Delta G|)
				\end{eqnarray}
				and the final posterior median~$\med\PMaRel$ (\tabref{tab_res_DQs}), we obtain the maximum change in relative angular position of the two components,
				\begin{equation}
					\DELangPos \lessapprox 0.311\UNITmuas.
				\end{equation}
				Comparison with \tabref{tab_data_used} reveals that this is more than two orders of magnitude smaller than the median AM measurement uncertainty, proving that the motions of the Earth and the CM of \miz\ can indeed be neglected.

			\begin{table*}
				\caption{Model parameters used by BASE.}\lbl{tab_pars}
				\centering
				\begin{tabular}{lllr@{, }lclcccc}
					\hline\hline
					&  &  & \multicolumn{2}{l}{Widest} &  & & \multicolumn{3}{c}{Data Types\,\tfm{b}}\\\cline{8-10}
					Symbol & Designation & Unit & \multicolumn{2}{l}{Prior Support} & Cyclic\,\tfm{a} & Prior Type & AM & RV & AM+RV\\\hline
					$\PMe$ & eccentricity & 1 & $[0$ & $1)$ &  & uniform & \chk & \chk & \chk\\
					$\PMf$ & orbital frequency & $\UNITdInvFlat$ & $(0$ & $1]$ &  & Jeffreys & \chk & \chk & \chk\\
					$\PMchi$ & mean anomaly at $\tmOne$ over $2\pi$ & 1 & $[0$ & $1)$ &\chk & uniform & \chk & \chk & \chk\\
					$\PMomegaP, \PMomegaii$ & argument of periapsis\,\tfm{c} & $\UNITrad$ & $[0$ & $2\pi)$ &\chk & uniform & \chk & \chk & \chk\\
					$\PMi$ & inclination & $\UNITrad$ & $[0$ & $\pi)$ &  & uniform & \chk &  & \chk\\
					$\PMOmega$ & position angle of the ascending node\tfm{e} & $\UNITrad$ & $[0$ & $2\pi)$ &\chk & uniform & \chk &  & \chk\\
					$\PMaRelApp$ & \bigCellAlt{.27\linewidth}{semi-major axis of orbit of secondary\\around primary over distance} & $\UNITmas$ & $[\EEi{-3}$ & $\EEi{-5}]$\,\tfm{f} &  & Jeffreys & \chk &  & \\
					$\PMpi$ & parallax & $\UNITarcsecTxt$ & $(0$ & $0.77]$\,\tfm{g} &  & Jeffreys &  &  & \chk\\
					$\PMV$ & RV offset & $\UNITmpsFlat$ & \multicolumn{2}{c}{---} &  & uniform &  & \chk & \chk\\
					$\PMK, \PMK_i$ & RV semi-amplitude\,\tfm{d} & $\UNITmpsFlat$ & \multicolumn{2}{c}{$\geq 0$} &  & mod. Jeffreys &  & \chk & \chk\\
					$\PMtauP$ & standard deviation of additional AM noise & $\UNITmas$ & \multicolumn{2}{c}{$\geq 0$} &  & mod. Jeffreys & \chk & & \chk\\
					$\PMsgmP$ & standard deviation of additional RV noise & $\UNITmpsFlat$ & \multicolumn{2}{c}{$\geq 0$} &  & mod. Jeffreys &  & \chk & \chk\\
					\hline
				\end{tabular}
				\tablefoot{
					\tablefoottext{a}{For cyclic parameters~$\param$, the indicated lower and upper bounds are treated as equivalent by BASE (see \secref{sec_obs_models}).}
					\tablefoottext{b}{The types of data for which each parameter is relevant. Abbreviations: AM (astrometry), RV (radial velocities).}
					\tablefoottext{c}{Normal mode employs $\PMomegaP$, while binary mode employs $\PMomegaii$.}
					\tablefoottext{d}{Normal mode employs $\PMK$, while binary mode employs $\PMKi$ and $\PMKii$.}
					\tablefoottext{e}{The widest prior range of $\PMOmega$ reduces to $[0, \pi)$ if no RV data are provided, in which case it cannot be determined whether a given node is ascending or descending; then, $\PMOmega$ is defined to be the position angle of the first node.}
					\tablefoottext{f}{Lower bound corresponds to AM measurement uncertainty of $1\UNITmuas$; upper bound according to wide-binary observations by \citet{380}.}
					\tablefoottext{g}{Interval includes trigonometric parallax of the nearest star, Proxima Centauri \citep{241}.}
				}
			\end{table*}

			\begin{table*}
				\caption{Quantities derived from model parameters.}\lbl{tab_DQs}
				\centering
				\begin{tabular}{r@{$\,\defby\,$}llllccccc}
					\hline\hline
					\multicolumn{2}{c}{} &  &  &  & \multicolumn{2}{c}{Mode\,\tfm{a}} & \multicolumn{3}{c}{Data Types\,\tfm{b}}\\\cline{6-7}\cline{8-10}
					\multicolumn{2}{l}{Definition} & Designation & Unit & Equations & N & B & AM & RV & AM+RV\\\hline
					$\PMP$ & $\PMf\inv$  & period & $\UNITd$ &  & \chk & \chk & \chk & \chk & \chk\\[.1cm]
					$\PMT$ & $\tmOne-\frac{\PMchi}{\PMf}$  & time of periapsis & $\UNITd$ & \eqrf{eq_def_chi} & \chk & \chk & \chk & \chk & \chk\\[.1cm]
					$\PMd$ & $\frac{1\UNITAU}{\PMpi}$  & distance & $\UNITpc$ &  &  & \chk &  &  & \chk\\[.1cm]
					$\PMrho$ & $\frac{\PMmC{2}}{\PMmC{1}} = \frac{\PMKi}{\PMKii}$  & binary mass ratio & $1$ & \eqrf{eq_def_CM}, \eqrf{eq_RV_amplitude_II} &  & \chk &  & \chk & \chk\\[.1cm]
					$\PMKLit$ & $\frac{\PMK}{\sqrt{1-\PMe^2}}$  & alternative RV semi-amplitude\,\tfm{c} & $\UNITmpsFlat$ &  & \chk & \chk &  & \chk & \chk\\[.1cm]
					$\PMmC{j}$ & $\frac{4\pi^2}{\gravConst}\,\frac{\PMK_{3-j}(\PMKi+\PMKii)^2}{\PMf(2\pi\sin\PMi)^3}$  & component mass & $\UNITMSun$ & \eqrf{eq_RV_amplitude_II}, \eqrf{eq_Kepler_III} &  & \chk &  &  & \chk\\[.1cm]
					$\PMaC{j}$ & $\frac{\PMK_j}{2\pi\PMf\sin\PMi}$  & \bigCellAlt{.25\linewidth}{semi-major axis of component's\\orbit around CM} & $\UNITAU$ & \eqrf{eq_RV_amplitude_II} &  & \chk &  &  & \chk\\[.1cm]
					$\PMaRel$ & $\PMaC{1}+\PMaC{2} = \frac{\PMKi+\PMKii}{2\pi\PMf\sin\PMi}$  & \bigCellAlt{.25\linewidth}{semi-major axis of secondary's\\orbit around primary} & $\UNITAU$ & \eqrf{eq_RV_amplitude_II} &  & \chk &  &  & \chk\\[.1cm]
					$\PMasini$ & $\frac{\PMK}{2\pi\PMf}$  & \bigCellAlt{.25\linewidth}{semi-major axis of stellar orbit\\times sine of inclination} & $\UNITAU$ & \eqrf{eq_RV_amplitude_II} & \chk &  &  & \chk & \\[.1cm]
					$\PMmmin$ & $\PMK(2\pi\gravConst\PMf)^{-\frac{1}{3}}(\PMmmin+\PMmS)^{\frac{2}{3}}$  & minimum planetary mass\,\tfm{d} & $\UNITMJup$ & \eqrf{eq_RV_amplitude_II} & \chk &  &  & \chk & \\[.1cm]
					$\PMaPMax$ & $\frac{\PMmS}{\PMmmin}\,\PMasini$  & \bigCellAlt{.25\linewidth}{maximum semi-major axis of\\planetary orbit} & $\UNITAU$ & \eqrf{eq_def_CM} & \chk &  &  & \chk & \\[.1cm]
					\hline
				\end{tabular}
				\tablefoot{
					\tablefoottext{a}{The modes in which each derived quantity appears. Abbreviations: N (normal), B (binary).}
					\tablefoottext{b}{The types of data for which each derived quantity is relevant. Abbreviations: AM (astrometry), RV (radial velocities).}
					\tablefoottext{c}{Refers to the star or any of the binary components, respectively.}
					\tablefoottext{d}{The implicit function of minimum planetary mass~$\PMmmin$ is solved numerically. The planet's minimum mass equals its real mass if the orbit is edge-on, \viz\ $\sin\PMi = 1$.}
				}
			\end{table*}

%
	\section{BASE -- Bayesian astrometric and spectroscopic exoplanet detection and characterisation tool}\lbl{sec_BASE}
%

		
		We have developed BASE, a computer program for the combined Bayesian analysis of AM and RV data according to \secref{sec_methodology}, for the following main reasons.
		\begin{itemize}
			\item A statistically well-founded, reliable tool was needed that was able to perform a complete Bayesian parameter and uncertainty estimation, along with model selection (only for planetary systems, not detailed in this article).
			
			\item We aimed to combine astrometry and Doppler-spectroscopy analyses.

			\item A possibility to include knowledge from earlier analyses was needed.

			\item Finding all relevant solutions across a multidimensional, high-volume parameter space~$\parSpace$ was required. A more detailed knowledge of the parameters than a best estimate and a confidence interval can provide is especially important when the data do not constrain the parameters well, \eg\ when only few data have been recorded or the \term{signal-to-noise} ratio is low (as can be the case for lightweight planets or young host stars).
		\end{itemize}

		
		BASE is a highly configurable command-line tool developed in Fortran~2008 and compiled with GFortran \citep{513}.
		Options can be used to control the program's behaviour and supply information such as the stellar mass or prior knowledge (see \secref{sec_prior_info}).
		Any option can be supplied in a configuration file and/or on the command line.

%
		\subsection{Prior knowledge}\lbl{sec_prior_info}
%


			Any model parameter can be assigned a prior probability density in one of three forms:

			\begin{itemize}
				\item a fixed value;

				\item a user-defined shape, \ie\ a set $\seti{(\param_i,\pPri{\param_i})}$;

				\item a range (either in combination with a specific shape, which is then clipped to the given range, or else using one of the standard prior shapes detailed in \secref{sec_priors}).
			\end{itemize}
			Any parameter for which no such option is present is assigned a default prior shape and range to pose the weakest possible restraints justified by the data, the model, and mathematical/physical considerations.

%
		\subsection{Physical systems and modes of operation}\lbl{sec_phys_systems}
%
		
			As mentioned above, BASE is capable of analysing data for two similar types of systems, whose physics have been described in \secref{sec_obs_models}. These are systems with
			
			\begin{enumerate}
				\item \textbf{one} observable component that may be accompanied by one or more unobserved bodies (generally referred to as planetary systems here for simplicity); in this \term{normal mode}, the number of companions can be set to a value ranging from zero to nine, or a list of such values, in which case several runs are conducted and the outcome is compared in terms of model selection;

				\item \textbf{two} observable, gravitationally bound components (referred to as binary stars); this mode of operation is called \term{binary mode}.
			\end{enumerate}

%
		\subsection{Types of data}\lbl{sec_data_types}
%

			Observational data of the following types can be treated by BASE:

			\begin{itemize}
				\item AM data, whose observable, in the case of binary targets, is the relative angular position of the two binary components. Each data record consists of a date, the angular position $(\POSalCosDel,\POSdel)$ or $(\POSrho,\POStheta)$ in a cartesian or polar coordinate system, and its standard uncertainty ellipse, given by $(\UNCERTELLa,\UNCERTELLb,\UNCERTELLphi)$;
				
				\item RV data, where each data record consists of a date and the observed stellar radial velocity as well as its uncertainty.
			\end{itemize}

%
		\subsection{Computational techniques}\lbl{sec_comput_techniques}
%

			In the following, we describe some computational techniques implemented in BASE that are relevant to the present work.
			
			\para{Improved exploration of parameter space.}
			To enhance the \term{mixing}, i.e. rapidness of exploration of the parameter space, of Markov chains produced by the Metropolis-Hastings (MH) algorithm (\secref{sec_posterior_inference}) and decrease their attraction by local posterior modes, the \term{parallel tempering (PT)} algorithm \citepeg{075} is employed one level above MH.
			Its function is to create in parallel $\nCh$ chains of length~$\nSamp$,
			$\left\{\vPar\indexCh\samp:\iSamp=1,\ldots,\nSamp;\iCh=1,\ldots,\nCh\right\}$
			each sampled by an independent MH procedure, where the \term{cold} chain $k=1$ uses an unmodified likelihood~$\likhood(\cdot)$, while the others, the \term{heated} chains, use as replacement $\likhood(\cdot)^{\tPar\indexCh}$ with the positive \term{tempering parameter}~$\tPar\indexCh < 1$.
			After $\nSwap$ samples of each chain, two chains $k\ge1$ and $k+1\le\nCh$ are randomly selected and their last links, denoted here by $\vPar\indxCh{k}$ and $\vPar\indxCh{k+1}$, are swapped with probability
			\begin{equation}
				\alpha_{\mathrm{swap}}(k,k+1) = \min\left(1,\left(\frac{\likhood(\vPar\indxCh{k+1})}{\likhood(\vPar\indxCh{k})}\right)^{\tPar_{k}}\left(\frac{\likhood(\vPar\indxCh{k})}{\likhood(\vPar\indxCh{k+1})}\right)^{\tPar_{k+1}}\right),\lbl{eq_swap_prob}
			\end{equation}
			which ensures that the distributions of both chains remain unchanged.
			
			This procedure allows states from the ``hotter'' chains, which explore parameter space more freely, to ``seep through'' to the cold chain without compromising its distribution. Conclusions are only drawn from the samples of the cold chain.
			
			In contrast to MCMC sampling, which is sequential in nature, the structure of PT allows one to exploit the multiprocessing facilities provided by many modern computing architectures. For this purpose, BASE uses the OpenMP API \citep{185} as implemented for GFortran by the GOMP project \citep{512}.

			\para{Assessing convergence.}
			As a matter of principle, it cannot be proven that a given Markov chain has converged to the posterior.
			However, convergence may be meaningfully defined as the degree to which the chain does not depend on its initial state~$\vPar\iniSamp$ any more. This can be determined on the basis of a set of independent chains -- in our case, the cold chains of $\nPT$ independent PT procedures -- started at different points in parameter space. These starting states should be defined such that for each of their components, they are \term{overdispersed} \wrt\ the corresponding marginal posteriors.
			
			Since the marginal posteriors are difficult to obtain before the actual sampling, BASE determines the starting states by repeatedly drawing, for each parameter, a set of $\nPT$ samples from the prior using \term{rejection sampling}; the repetition is stopped as soon as the sample variance exceeds the corresponding prior variance, which yields a set of starting states overdispersed \wrt\ the prior. Assuming that the prior variance exceeds the marginal-posterior variance, the overdispersion requirement is met.

			Such a test, using the \term{potential scale reduction (PSR)} or Gelman-Rubin statistic, was proposed by \citet{514} and was later refined and corrected by \citet{322}. It is repeatedly carried out during sampling and, in the case of a positive result, sampling is stopped before the user-defined maximum runtime and/or number of samples have been reached. It may also sometimes be useful to abort sampling manually, which can be done at any time.
			
			The statistic is calculated with respect to each parameter $\param$ separately, using the post-burn-in samples%
			\footnote{Because the burn-in length~$\nSampBurn$ cannot be determined in advance, BASE considers a fixed but configurable fraction of samples to belong to the burn-in phase at any time.}
			$\left\{\param\indexPT\samp:\iSamp=\nSampBurn+1,\ldots,\nSamp;\iPT=1,\ldots,\nPT\right\}$ of $\param$ provided by the $\nPT$ independent chains.
			It compares the actual variances of $\param$ within the chains built up thus far to an estimate of the marginal-posterior (\ie\ target) variance of $\param$, thus estimating how closely the chains have approached convergence.

			The PSR~$\PSR$ is defined as
			\begin{equation}
				\PSRSq \defby \frac{\PSRDOF+3}{\PSRDOF+1}\,\frac{\PSRV}{\PSRW},
			\end{equation}
			where $\PSRV$ is an estimate of the marginal-posterior variance, $\PSRDOF$ the estimated number of degrees of freedom underlying the calculation of $\PSRV$, and $\PSRW$ the mean within-sequence variance. The quantities are defined as
			\begin{eqnarray}
				\PSRV &\defby& \frac{\nu-1}{\nu}\PSRW+\frac{\nPT+1}{\nPT\nu}\PSRB\\
				\PSRW &\defby& \frac{1}{\nPT(\nu-1)}\sum_{\iPT=1}^\nPT \sum_{\iSamp=\nSampBurn+1}^\nSamp \left(\param\indexPT\samp-\overline{\param\indexPT}\right)^2\\
				\PSRB &\defby& \frac{\nu}{\nPT-1} \sum_{\iPT=1}^\nPT \left(\overline{\param\indexPT}-\overline{\param}\right)^2,
			\end{eqnarray}
			where $\PSRB$ is the between-sequence variance and $\nu \defby \nSamp-\nSampBurn$; a horizontal bar denotes the mean taken over the set of samples obtained by varying the omitted indexes.
			
			Owing to the initial overdispersion of starting states, $\PSRV$ overestimates the marginal-posterior variance in the beginning and subsequently decreases. Furthermore, while the chains are still exploring new areas of parameter space, $\PSRW$ underestimates the marginal-posterior variance and increases. Thus, as convergence to the posterior is accomplished, $\PSR \searrow 1$. Therefore, we can assume convergence as soon as the PSR has fallen below a threshold~$\PSRthr \ge 1$. If BASE is run with $\PSRthrSq \defby 1$, sampling is continued up to the user-defined length or duration, respectively.

			For cyclic parameters (\secref{sec_obs_models}), the default lower and upper prior bounds are equivalent, which needs to be taken into account when calculating the PSR. Thus, BASE uses the modified definition of the PSR by \citet{023} for these parameters.

%
	\section{Target and data}\lbl{sec_target+data}
%

		\begin{table}
			\caption{Basic physical properties of \miz.}\lbl{tab_miz_basic_props}
			\centering
			\begin{tabular}{lr@{$\,\pm\,$}l@{ }ll}
				\hline\hline
				Property & \multicolumn{3}{l}{Value} & Reference\\\hline
				Type & \multicolumn{3}{l}{SB\,II} & 1\\
				Spectral types & \multicolumn{3}{l}{$2\times$\,A2\,V} & 2\\
				$M_\mathrm{V}$ & 2.27 & 0.07 & mag & 2\\
				$M_{\mathrm{bol}}$ & 0.91 & 0.07 & mag & 3\\
				$L$ & 33.3 & 2.1 & $L_{\sun}$ & 3\\
				$R$ & 2.4 & 0.1 & $R_{\sun}$ & 3\\
				$\varpi$\,\tfm{a} & 39.4 & 0.3 & $\UNITmas$ & 3\\
				$\propMotAl$ & 119.01 & 1.49 & $\UNITmas\,\UNITyr\inv$ & 4\\
				$\propMotDel$ & -25.97 & 1.65 & $\UNITmas\,\UNITyr\inv$ & 4\\
				\hline
			\end{tabular}
			\tablefoot{
				\tablefoottext{a}{Different values of the parallax have been estimated in this work (\tabref{tab_res_pars}).}
			}
			\tablebib{(1)~\citet{458}; (2)~\citet{459}; (3)~\citet{260}; (4)~\citet{456}}
		\end{table}

		\begin{table*}
			\caption{Published data for \miz\ used in this work.}\lbl{tab_data_used}
			\centering
			\begin{tabular}{llllll@{\,}l@{\,/\,}l@{\,}ll}\hline\hline
				Data type & Instrument & Observatory & Year & No. records & \multicolumn{4}{l}{Median uncertainty} & Reference\\\hline
				Radial velocities & Coudé spectrograph & Haute Provence & 1961 & $2\times 17$ & 2.050 & \multicolumn{3}{l}{$\UNITkmpsFlat$\,\tfm{a}} & 1\\
				Angular positions & Mark~III interferometer & Mount Wilson & 1995 & 28 & 0.040 & $\UNITmas$ & 0.345 & $\UNITmas$\,\tfm{b} & 2\\
				Angular positions & NPOI interferometer & Lowell & 1998 & 25 & 0.042 & $\UNITmas$ & 0.137 & $\UNITmas$\,\tfm{b} & 3\\
				\hline
			\end{tabular}
			\tablefoot{
				\tablefoottext{a}{Uncertainties are missing in the original publication, but have been estimated in this work (\secref{sec_prep_data}).}
				\tablefoottext{b}{Uncertainties refer to the semi-minor and -major axes of the standard uncertainty ellipses, respectively (\secref{sec_freq_inference}).}
			}
			\tablebib{(1)~\citet{290}; (2)~\citet{280}; (3)~\citet{260}}
		\end{table*}


		\miz\ (\mizAltNames), the first spectroscopic binary discovered, is of double-lined type \citep[\SBII;][]{458}. Its basic physical properties are summarised in \tabref{tab_miz_basic_props}. Together with the spectroscopic binary \object{\mizb}, it forms the \object{Mizar} quadruple system, seen from Earth as a visual binary with the components separated by about $14\farcs4$.
		Mizar is the first double star discovered by a telescope and also the first one to be imaged photographically \citep{460}.
		
		At an apparent angular separation of about $11\farcm8$, or $74\pm39\UNITkAU$ spatial distance, Mizar is accompanied by Alcor, which has recently turned out to be a spectroscopic binary itself \citep{461}. Mizar and Alcor, also known as the ``Horse and Rider'', form an easy naked-eye double star, while it is still a matter of debate whether or not they constitute a physically bound sextuplet.

		
		Published data used in this article are displayed in \figsrefrange{fig_res_da_mo_re_all}{fig_res_da_mo_re_all_fold} along with the model functions determined by BASE and their properties are summarised in \tabref{tab_data_used}. Using a Coud\'e spectrograph at the 1.93-m telescope at Observatoire Haute Provence, \citet{290} obtained 17 optical photographic spectra of the light of \miz\ combined with that of electric arcs or sparks between iron electrodes. For each of the 13 individual stellar lines identified in the spectra, one intermediate radial-velocity value per binary component was obtained by comparison with a set of reference lines of iron. Finally, a set of 17 pairs of RVs for the two components was calculated as arithmetic means of the corresponding intermediate values. The RV measurement uncertainty is not given by \citet{290} and was therefore estimated in the course of the present work, as described in \secref{sec_results}.

		High-precision AM data of \miz\ were first obtained by \citet{280} using the Mark~III optical interferometer on Mount Wilson, California \citep{296}, with baseline lengths between 3 and 31 m. It measured the squared \term{visibilities} and their uncertainties at positions sampled over the aperture plane due to Earth's rotation. The visibilities can be modelled as a function of the diameters, magnitude differences, and relative angular positions $(\POSrho,\POStheta)$ of the binary components, \citet{284}. These authors also describe a procedure to derive one angular position for each night of observation from a corresponding set of visibilities, which was adopted by \citet{280} to obtain initial estimates of the orbital parameters. These positional data are also relevant for the present work. \citet{280} also performed a direct fit to the squared visibilities to derive final estimates of the component diameters and orbital parameters.
		
		Later, a descendant instrument, the Navy Prototype Optical Interferometer (NPOI) at Lowell observatory, Arizona \citep{297}, was used by \citet{260} to obtain more accurate results using three siderostats, \viz\ three baselines at a time. This allowed a better calibration using the \term{closure phase},%
		\footnote{The \term{closure phase}~$\clph$ is the phase of the product of three visibilities, each pertaining to a different baseline.}
		which is independent of atmospheric turbulence. Similarly as before, \citet{260} separately fitted binary orbits directly to the visibility data and also to the positional angles derived for each night, concluding that the respective results agree well with each other and that those parameters in common with spectroscopic analyses are compatible with \citet{290}; they also performed a fit to both AM and RV data to obtain their final parameter estimates.
		
		While \citet{260} did not include the older, less accurate Mark~III data in their analysis, we present a combined treatment of all published data, \ie\ the AM positions of \citet{280} and \citet{260} along with the RV data of \citet{290}.

%
	\section{Analysis and results}\lbl{sec_results}
%
	

		\begin{table}
			\caption{Analysis passes carried out sequentially.}\lbl{tab_runs}
			\centering
			\begin{tabular}{lll}
				\hline\hline
				Pass & Description & Data\\
				\hline
				A & First constraints on RV~parameters & RV\\
				B & Combining all data & All\\
				C & Selecting frequency~$\PMf$ & All\\
				D & Selecting $\PMomegaii$ and $\PMOmega$ & All\\
				E & Refining results & All\\
				\hline
			\end{tabular}
		\end{table}

		To illustrate the features of BASE and demonstrate its validity, we used the tool to analyse all published data of \miz. This section details the steps taken in and the results of our analysis. Its general goal was, given uninformative prior knowledge, to search the parameter space as comprehensively as possible to find and characterise the a posteriori most probable solution together with its uncertainty and other characteristics. For reasons detailed below, once the RV data had been prepared, several runs of BASE (\tabref{tab_runs}) were manually conducted, each using priors derived from the previous pass, with relatively uninformative priors in the first step.
		
		In this analysis, our approach was to regard all nominal measurement uncertainties as accurate by setting parameters characterising additional noise in AM ($\PMtauP$) as well as RV ($\PMsgmP$) data to zero.
		
		Astrometric data alone allow one to constrain neither the RV offset~$\PMV$ nor the amplitudes~$\PMKi,\PMKii$ but only the sum~$\PMKi+\PMKii$ (\secref{sec_observables}). By contrast, these parameters can be constrained using spectroscopic data alone, or AM and RV data both. However, the AM data reduce the relative weight of the spectroscopic data and thus make the determination of these parameters harder. We found in the course of this work that by an iterative approach, starting with a first pass using only RV data, this difficulty could be resolved.
		
		In all passes, BASE was configured to build up eight parallel chains, including the cold chain, using the parallel-tempering technique (detailed in \secref{sec_comput_techniques}). $\EEi{8}$ posterior samples were collected, the first 10\% of which were assumed to be burn-in samples. In the final pass, two PT procedures were employed to enable a test of convergence to the posterior, which reduced the number of samples collected with unchanged memory requirements to $\EE{5}{7}$. Convergence was not assessed in earlier passes, because convergence was difficult to reach as long as several distinct solutions existed within the prior support.

%
		\subsection{Preparation of RV data}\lbl{sec_prep_data}
%

			Assuming appropriate measurement uncertainties is a prerequisite for the proper relative weighting of the different data types when combining them. Because these uncertainties are not quoted by \citet{290} for the spectroscopic data, we estimated them according to the following method.
			
			First, we assumed that each RV datum~$\RV_i$ is the sum of the model value~$\mdlFctRV{\vParTrue}{\tm_i}$ and an error~$e_i$, where $\vParTrue$ are the true parameter values and the errors are independent and identically Gaussian-distributed with an unknown standard deviation~$\measUnc$. Furthermore, we assumed that the model parameters found in a particular analysis are identical with $\vParTrue$. It follows that the uncertainty~$\measUnc$ can be estimated as the sample standard deviation of the set of residuals~$\seti{\RV_i-\mdlFctRV{\vParTrue}{\tm_i}}$.
			
			Thus, based on the sample standard deviation of the best-fit residuals of \citet{290}, \viz\ $2.13\UNITkmpsFlat$, we initially assumed a conservative value of $2.50\UNITkmpsFlat$ for all data. To quantify the measurement uncertainties based on our own inference, we then conducted a preliminary analysis similar to \run{A} described in the next subsection, from which we finally inferred $\measUnc = 2.05\UNITkmpsFlat$. In addition, we took a more correct alternative approach to estimating the RV uncertainties by assuming a relatively low value of $\measUnc=2.00\UNITkmpsFlat$ and allowing for higher noise via $\PMsgmP$, which led to an estimate $\sqrt{\measUnc^2+\PMsgmP^2}$ deviating by less than 1.7\% from our previous estimate.

%
		\subsection{\Run{A}: first constraints on RV~parameters}\lbl{sec_ana_rA}
%
	
			\begin{table}
				\caption{\Run{A}: initial prior ranges.}\lbl{tab_ana_rA_priors}
				\centering
				\begin{tabular}{r@{.}lr@{.}lr@{.}lr@{.}lr@{.}lr@{.}lr@{.}l}
					\hline\hline
					\multicolumn{2}{l}{$\PMe$} & \multicolumn{2}{l}{$\PMf$\,\tfm{a}} & \multicolumn{2}{l}{$\PMchi$} & \multicolumn{2}{l}{$\PMomegaii$} & \multicolumn{2}{l}{$\PMV$\,\tfm{b}} & \multicolumn{2}{l}{$\PMKi$\,\tfm{c}} & \multicolumn{2}{l}{$\PMKii$\,\tfm{c}}\\
					\multicolumn{2}{l}{} & \multicolumn{2}{l}{$(\UNITdInvFlat)$} & \multicolumn{2}{l}{} & \multicolumn{2}{l}{$(\UNITrad)$} & \multicolumn{2}{l}{$(\UNITkmpsFlat)$} & \multicolumn{2}{l}{$(\UNITkmpsFlat)$} & \multicolumn{2}{l}{$(\UNITkmpsFlat)$} \\\hline
					\multicolumn{2}{l}{$0$} & $2$ & $\EE{7379}{-6}$ & \multicolumn{2}{l}{$0$} & \multicolumn{2}{l}{$0$} & $-65$ & $25$ & \multicolumn{2}{l}{$0$} & \multicolumn{2}{l}{$0$}\\
					\multicolumn{2}{l}{$1$} & \multicolumn{2}{l}{$1$} & \multicolumn{2}{l}{$1$} & \multicolumn{2}{l}{$2\pi$} & $56$ & $15$ & $69$ & $50$ & $68$ & $85$\\
					\hline
				\end{tabular}
				\tablefoot{
					\tablefoottext{a}{The prior bounds of $\PMf$ correspond to a period range between $1\UNITd$ and $1\UNITyr$.}
					\tablefoottext{b}{The bounds of $\PMV$ are given by the section of the RV ranges measured for primary and secondary, with the latter extended by one corresponding measurement uncertainty on both sides.}
					\tablefoottext{c}{The upper bounds of $\PMKi$ and $\PMKii$ are given by half the measured RV span of the corresponding component, plus one measurement uncertainty.}
				}
			\end{table}
			
			\begin{figure}
				\centering
				\inclGrII{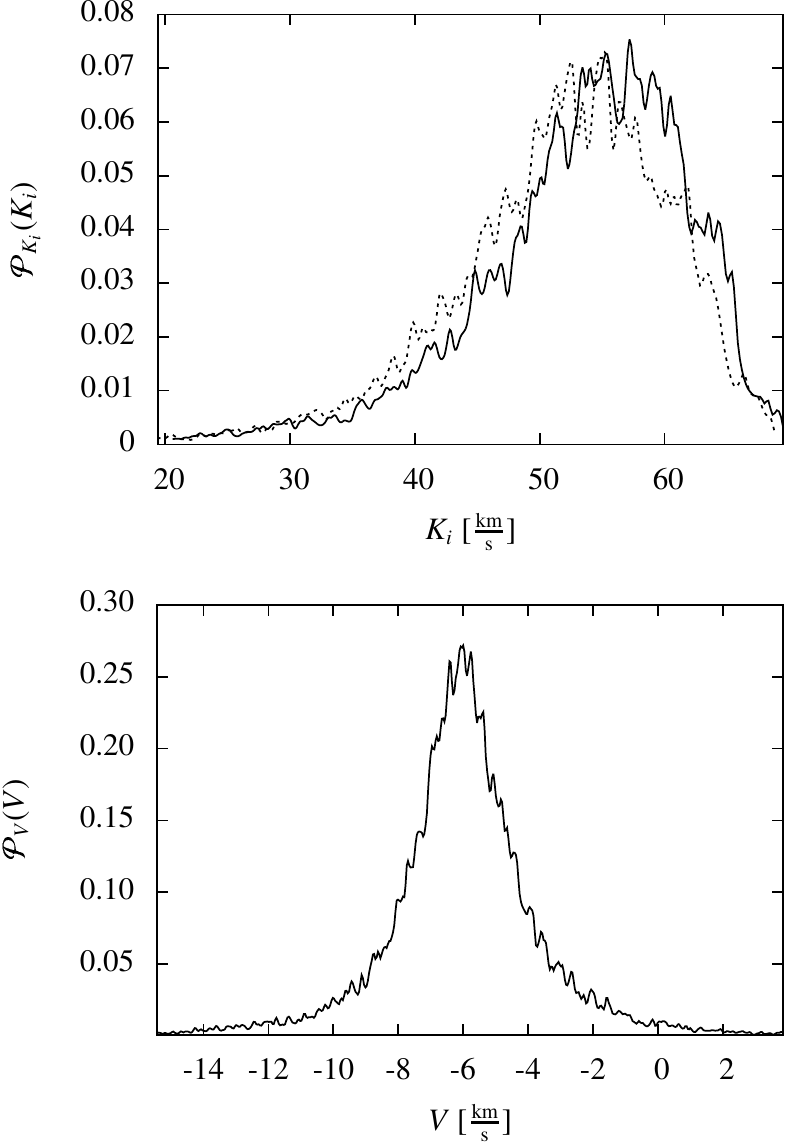}
				\caption{\Run{A}: Marginal posteriors of RV amplitudes $\PMKi$ (top, solid line), $\PMKii$ (top, dashed line) and offset~$\PMV$ (bottom), all plotted over the approximate range of the corresponding 99\% HPDI.}\lbl{fig_res_rA_mapo_K1_K2_V}
			\end{figure}

			To facilitate the determination of the RV parameters $\PMKi,\PMKii$ and $\PMV$, a first pass using only RV data was carried out, as mentioned above. It used uninformative priors (\secref{sec_priors} and \tabref{tab_pars}), with bounds listed in \tabref{tab_ana_rA_priors}.
			
			\Figref{fig_res_rA_mapo_K1_K2_V} shows the resulting marginal posteriors (see \secref{sec_posterior_inference}) of the RV offset~$\PMV$ and the amplitudes~$\PMKi,\PMKii$, with the abscissae approximately corresponding to the 99\% HPDIs. These marginal posteriors represent much tighter constraints on the parameters than the corresponding priors do.

%
		\subsection{\Run{B}: combining all data}\lbl{sec_ana_rB}
%
		
			\begin{figure}
				\centering
				\inclGrII{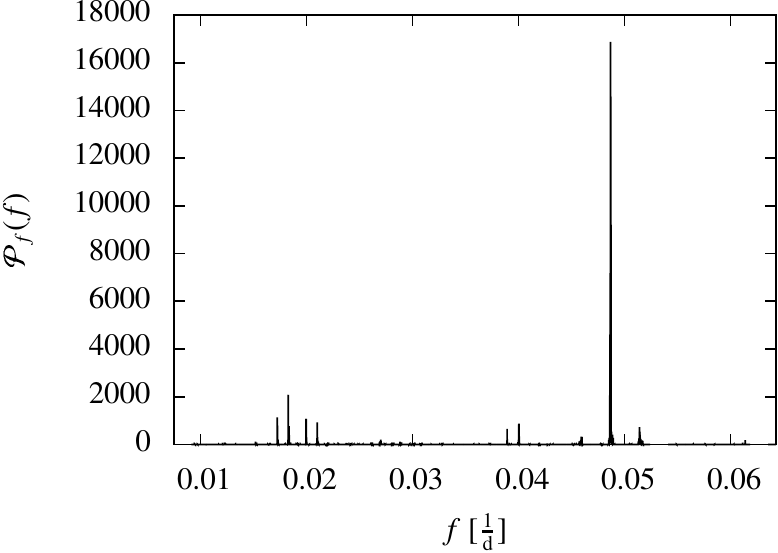}
				\caption{\Run{B}: Marginal posterior of orbital frequency~$\PMf$, plotted over the range of the corresponding 99\% HPDI. This is a Bayesian analogon to the frequentist periodogram.}\lbl{fig_res_rB_mapo_f}
			\end{figure}
	
			\begin{table}
				\caption{\Run{B}: prior ranges for additional AM parameters.}\lbl{tab_ana_rB_priors}
				\centering
				\begin{tabular}{r@{.}lr@{.}lr@{.}l}
					\hline\hline
					\multicolumn{2}{l}{$\PMi$} & \multicolumn{2}{l}{$\PMOmega$} & \multicolumn{2}{l}{$\PMpi$\,\tfm{a}}\\
					\multicolumn{2}{l}{$(\UNITrad)$} & \multicolumn{2}{l}{$(\UNITrad)$} & \multicolumn{2}{l}{$(\UNITarcsecTxt)$}\\\hline
					\multicolumn{2}{l}{$0$} & \multicolumn{2}{l}{$0$} & \multicolumn{2}{l}{$0$}\\
					\multicolumn{2}{l}{$\pi$} & \multicolumn{2}{l}{$2\pi$} & \multicolumn{2}{l}{$0.77$}\\
					\hline
				\end{tabular}
				\tablefoot{
					\tablefoottext{a}{Interval includes trigonometric parallax of nearest star, Proxima Centauri \citep{241}.}
				}
			\end{table}
			
			Providing as new priors the marginal posteriors of all RV parameters \vParRVCompoTxt\ from \run{A}, constrained to the corresponding 99\% HPDIs~$\HPDIArg{99\%}$, and again using uninformative priors on the additional parameters (\tabref{tab_ana_rB_priors}), the AM data were added and BASE was run again. Using the resulting posterior samples, BASE was additionally invoked in periodogram mode to refine the kernel window width for the marginal posterior of $\PMf$ as described in \secref{sec_posterior_inference}. (This refinement was not used in \run{A} in order not to constrain the frequency too tightly before adding the AM data, because the combined data are expected to correspond to a marginally differing frequency.)

			The resulting marginal posterior (\figref{fig_res_rB_mapo_f}) exhibits a very strong mode around $0.04869\UNITdInvFlat$, whose height is 8.1 times that of the next lower peak. Over the range of its mode, the marginal posterior contains a total probability of 45.0\%. Most other parameters in this stage have very broad and/or multimodal marginal posteriors, hinting at different solutions still probable within the prior support.

%
		\subsection{\Run{C}: selecting the frequency~$\PMf$}\lbl{sec_ana_rC}
%
		
			\begin{figure}
				\centering
				\inclGrII{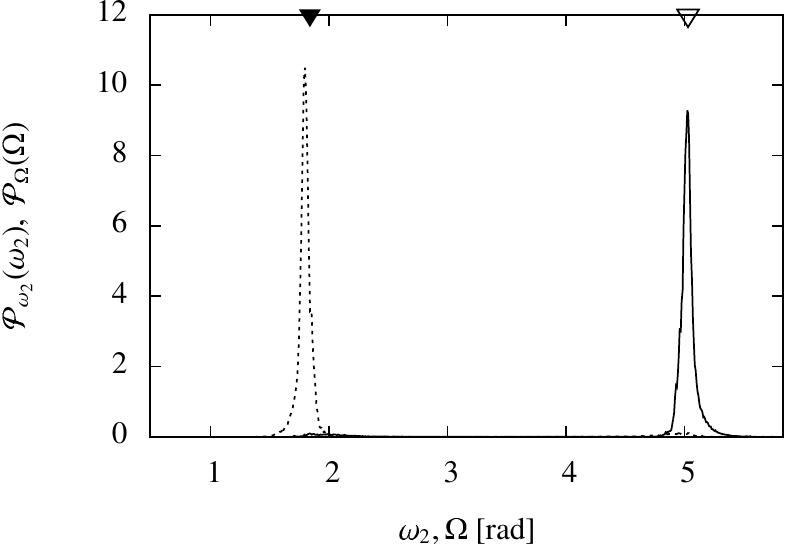}
				\caption{\Run{C}: Marginal posteriors of the argument of periapsis~$\PMomegaii$ (solid line) and position angle of the ascending node~$\PMOmega$ (dashed line), plotted over the approximate range of the corresponding 99\% HPDIs. Triangles indicate the positions of small local maxima located approximately $\pm\pi$ from the corresponding marginal modes.}\lbl{fig_res_rC_mapo_omega_2_Omega}
			\end{figure}
		
			For the next pass, the prior of $\PMf$ was provided by the marginal posterior of \run{B} (\figref{fig_res_rB_mapo_f}), constrained to the range of the marginal mode. For all other parameters, priors were identical with the previous marginal posteriors, constrained to $\HPDIArg{99\%}$.
			
			With the frequency thus restrained, the pass produced unimodal marginal posteriors in all parameters except for $\PMomegaii$ and $\PMOmega$.
			
			The marginal posteriors of the argument of periapsis~$\PMomegaii$ and the position angle of the ascending node~$\PMOmega$ exhibited small local maxima located at a distance of $-1.02\pi$ and $+1.03\pi$ from their marginal modes, respectively (indicated by triangles in \figref{fig_res_rC_mapo_omega_2_Omega}). 
			This can be explained by the fact that the Thiele-Innes constants (\eqsrefrange{eq_rot_matrix_A}{eq_rot_matrix_G}), which appear in the AM model, contain products of $\sin(\cdot)$ and/or $\cos(\cdot)$ functions with $\PMomegaii$ and $\PMOmega$ as arguments. These products retain their values when $\PMomegaii$ and $\PMOmega$ are both shifted by $\pi$ -- with opposite signs, since the marginal modes of these angles lie in different halves of the interval~$[0,2\pi)$. Consequently, the AM model function is invariant \wrt\ such shifts. (Thus, when analysing only AM data, it cannot be determined which node is ascending, hence $\PMOmega$ is defined only over the interval~$[0,\pi)$ and refers to the first node.) Owing to the combination with RV data, which independently constrain $\PMomegaii$ (\eqref{eq_RV_S_five}), the ambiguity is strongly reduced, as illustrated by the very small subpeaks in \figref{fig_res_rC_mapo_omega_2_Omega} -- though it is not completely resolved.

			As another result from this pass, the high correlation coefficient of the two angles, $\corrCoeff{\PMomegaii}{\PMOmega}=-0.62$, expresses the strong negative linear relationship between them introduced by the possibility of a contrarious change.

%
		\subsection{Passes D and E: selecting $\PMomegaii$ and $\PMOmega$ and refining results}\lbl{sec_ana_rD}
%
		
			In \run{D}, the ambiguity in $\PMomegaii$ and $\PMOmega$ was resolved by selecting the range around their marginal modes via the new priors. For all other parameters, priors were again identical to the previous marginal posteriors, constrained to $\HPDIArg{99\%}$. All resulting marginal posteriors turned out to be unimodal, corresponding to a single solution as opposed to several clearly distinct orbital solutions.
		
			As a final step, \run{E} was conducted to refine the results by using all previous marginal posteriors, constrained to $\HPDIArg{99\%}$, as new priors, thus confining the parameter space a priori to the most probable solution.


			\begin{figure*}
				\centering
				\inclGrII{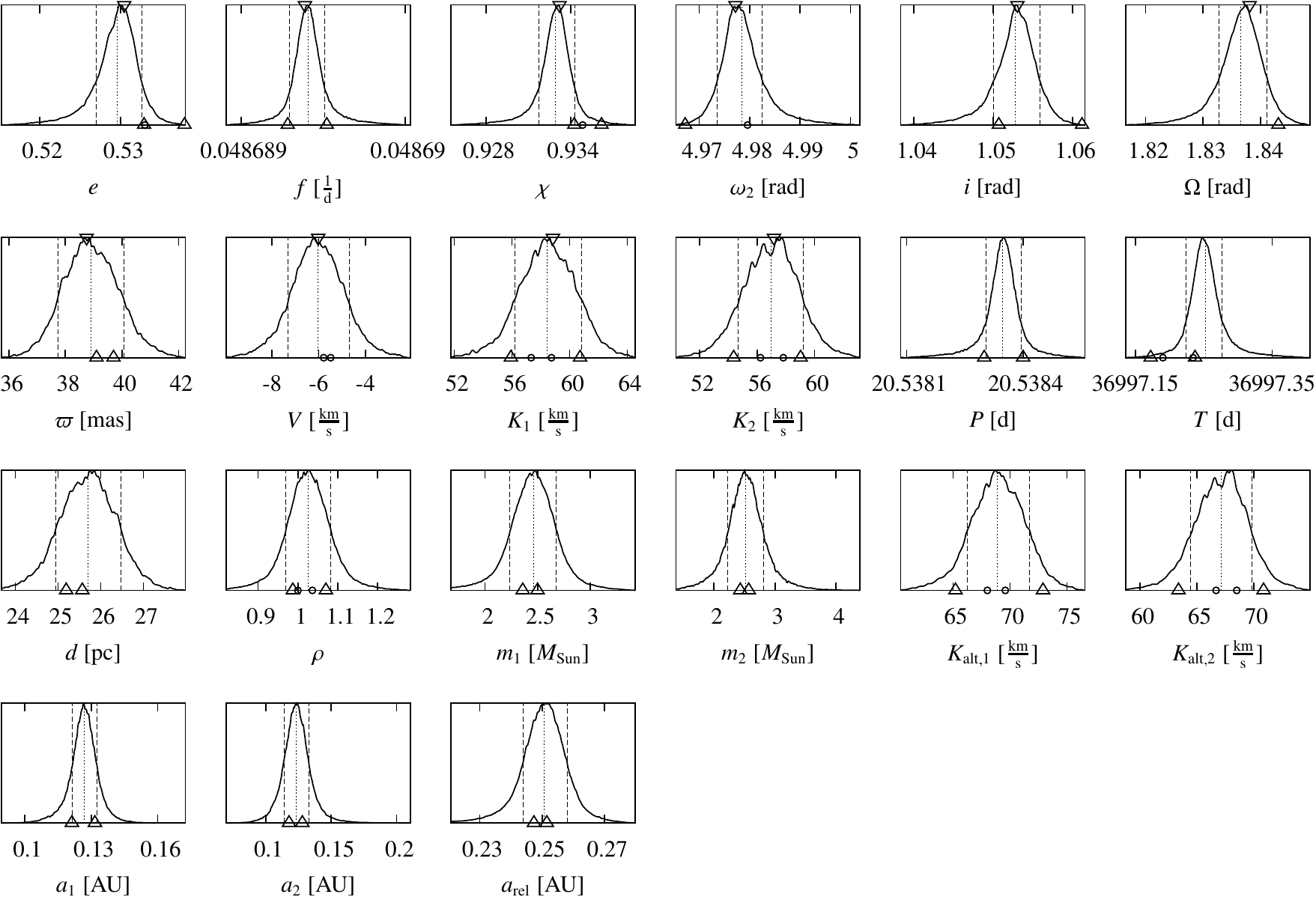}
				\caption{Marginal posteriors of parameters $\vParAllCompo$ (see also \tabref{tab_pars}) and derived quantities $\vDQAllCompo$ (\tabref{tab_DQs}) with marginal-posterior medians (dotted line) and 68.27\% HPDIs (dashed lines), from the final pass. Abscissae ranges are identical with the corresponding 95\% HPDIs. Ordinate values were omitted but follow from normalisation. Open triangles on the upper abscissae indicate the MAP estimates. Open circles on the lower abscissae bound the confidence intervals given by or derived from \citet{290}, while open triangles refer to the intervals of \citet{260}. Some of these literature estimates are not plotted because they are outside the abscissa range; for $\PMf$ and $\PMchi$, no literature uncertainty estimate is available. For derivations and numerical values, see \tabsrefii{tab_res_pars}{tab_res_DQs}.}\lbl{fig_res_mapo}
			\end{figure*}
					
			\begin{figure*}
				\centering
				\inclGrPNG{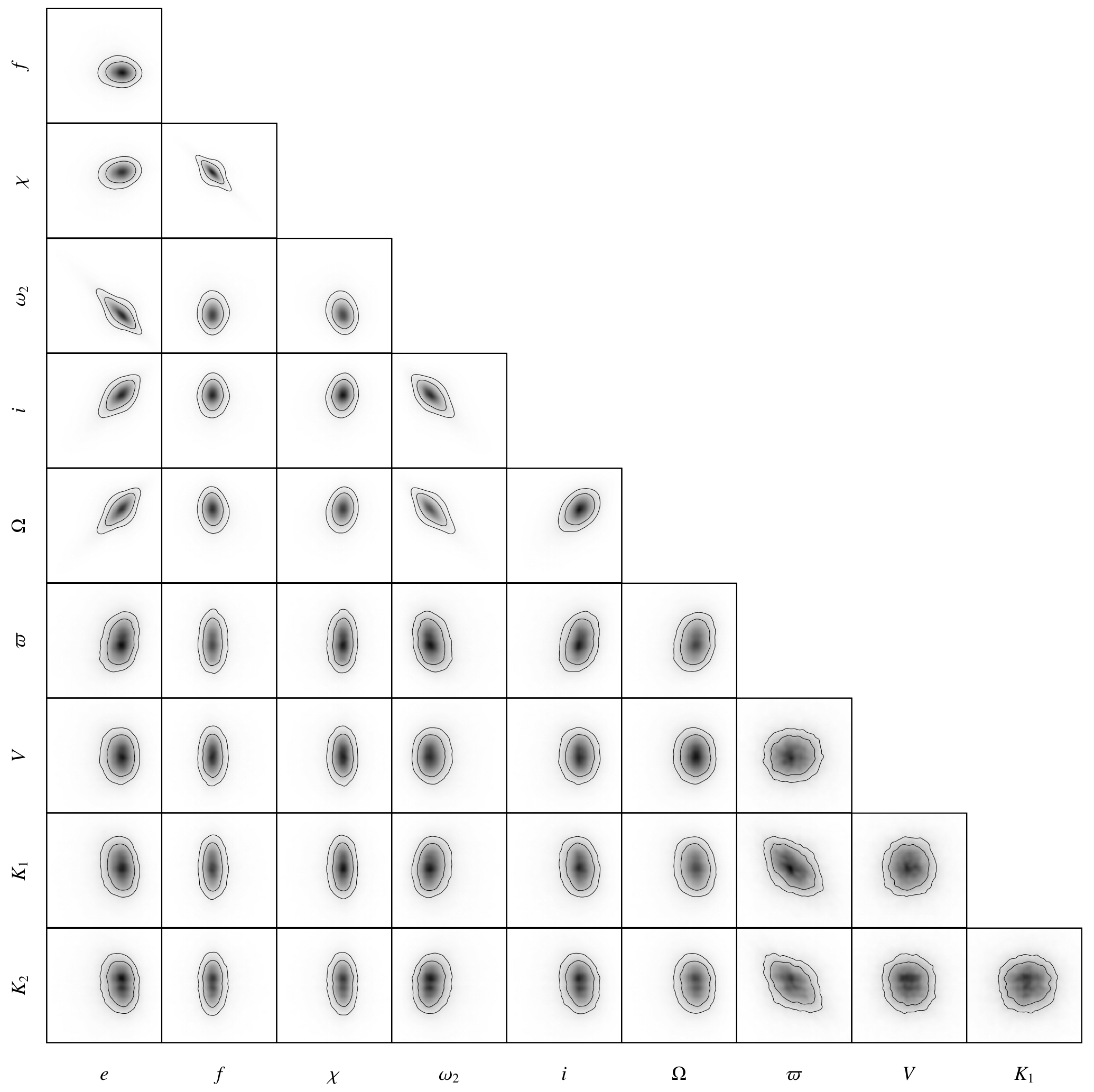}
				\caption{Two-parameter joint marginal posterior densities~$\jointmapo{i}{j}{\cdot}{\cdot}$ from \run{E}. The figure consists of 45 sub-plots, one for each combination of two parameters. Black denotes highest density. The inner and outer contours contain 50\% and 64.86\% probability, respectively. All plots aligned in one column share the same abscissa, denoted on the bottom; all plots aligned in one row share the same ordinate, denoted on the left. All abscissae and ordinates are displayed over the corresponding 95\% HPDIs (\tabref{tab_res_pars}), such that the total probability content of each plot is between 90\% and 95\%.
				}\lbl{fig_res_JMP}
			\end{figure*}


			\begin{figure*}
				\centering
				\inclGrII{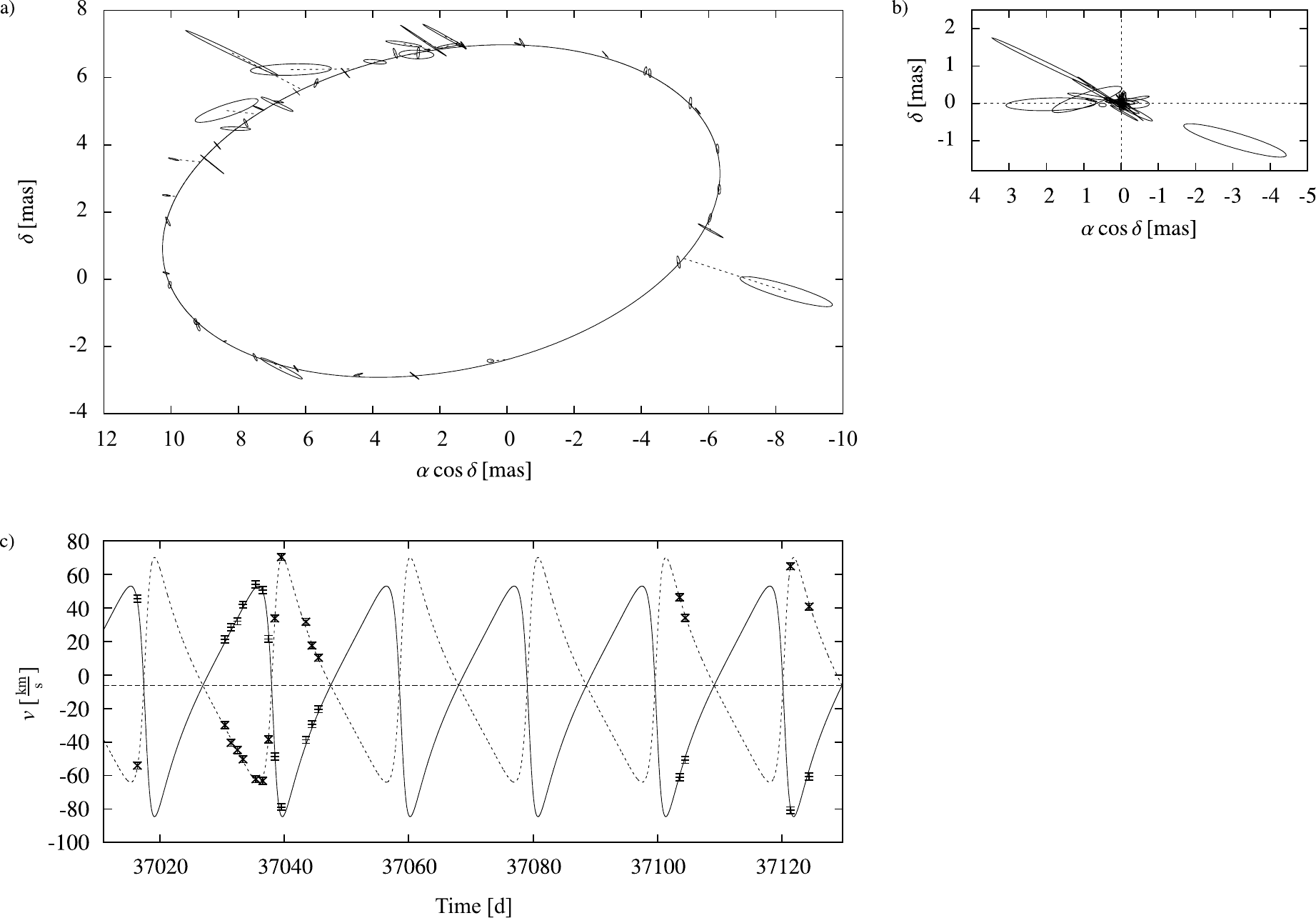}
				\caption{AM and RV data and models calculated with the MAP parameter estimates (\tabref{tab_res_pars}). a) AM data (uncertainty ellipses, solid lines), model (large ellipse, solid line) and residual vectors (dashed lines). b) Residual AM error ellipses. c) RV data and model for primary (normal error bars, solid line) and secondary (hourglass-shaped error bars, dotted line). The horizontal dashed line indicates the RV offset~$\PMV$. RV residuals are presented in \figref{fig_res_da_mo_re_all_fold}.}\lbl{fig_res_da_mo_re_all}
			\end{figure*}

					\begin{figure}
						\centering
						\hskip-1cm
						\inclGrII{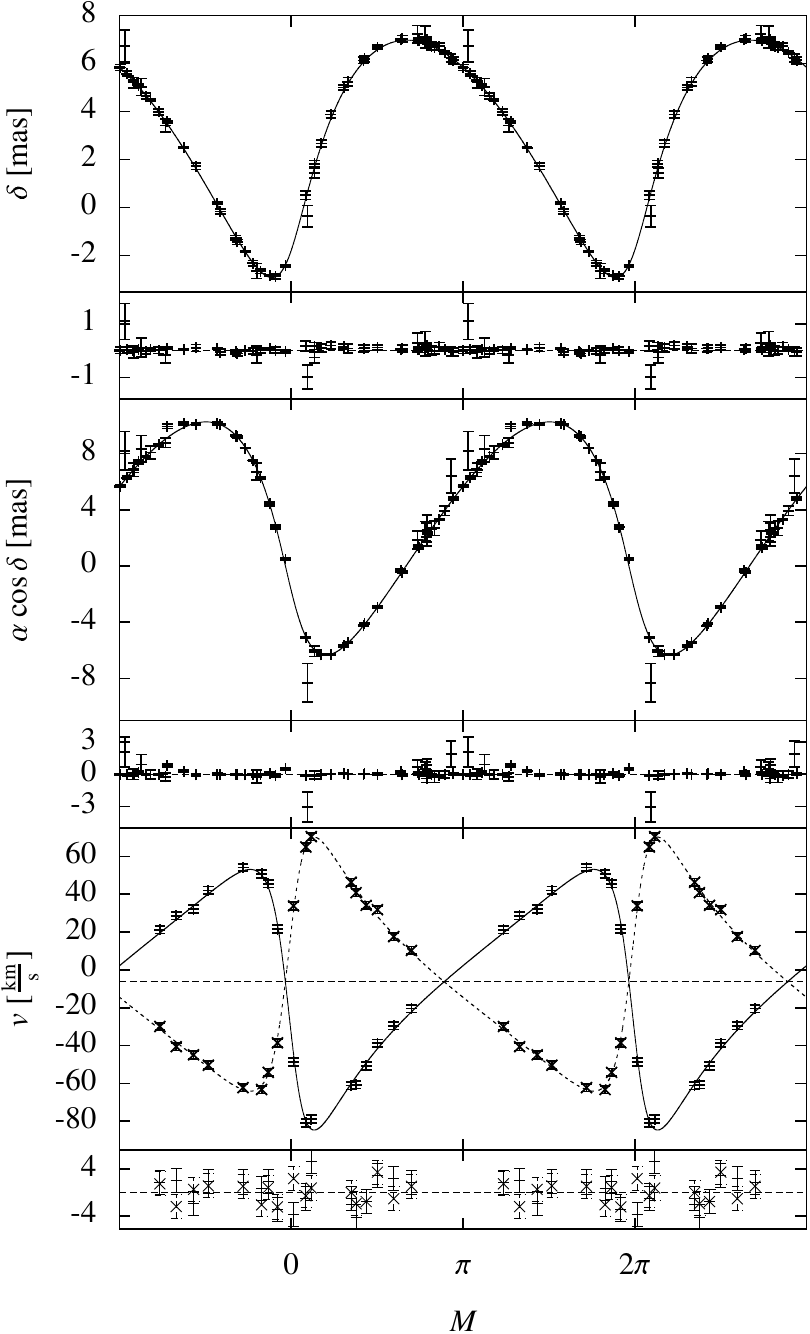}
						\caption{All data types: from top to bottom: AM data and model ($\POSdel$~coordinate); AM residuals~$\Delta\POSdel$; AM data and model ($\POSalCosDel$~coordinate); AM residuals~$\Delta(\POSalCosDel)$; RV data and model $\RV$ (primary: solid line, secondary: dotted line, RV offset~$\PMV$: horizontal dashed line); RV residuals~$\Delta\RV$ (primary: normal error bars, secondary: dashed error bars, model: dashed line). AM error bars are defined as the sides of the smallest rectangle orientated along the coordinate axes and containing the respective error ellipse. Abscissa values are mean anomalies~$\MA$ (\eqref{eq_Keplers_eq_chi}), \ie\ times \term{folded} \wrt\ the MAP estimates of the time of periapsis~$\PMT$ and period~$\PMP$.}\lbl{fig_res_da_mo_re_all_fold}
					\end{figure}

			\begin{figure}
				\centering
				\inclGrII{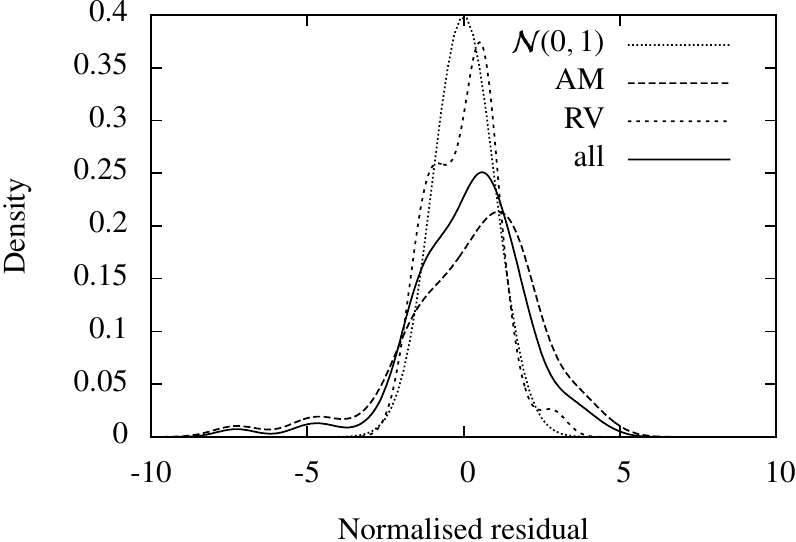}
				\caption{Distribution of normalised residuals of AM data (long-dashed line), RV data (short-dashed) and all data (solid), along with the standard normal distribution~$\stdNormDistr$ (dotted). For definitions of the normalised residuals, see \eqsrefrange{eq_def_norm_resid_AM}{eq_chi_sq_RV_simple}.}\lbl{fig_res_distr_norm_resid_all}
			\end{figure}

			\paragraph{(Joint) marginal posteriors and correlations.}
			The final marginal posteriors of all model parameters, plotted over the corresponding 95\%~HPDIs, are shown in \figref{fig_res_mapo}, along with the medians, 68.27\% HPDIs (corresponding in probability content to the frequentist 1-$\sigma$ confidence intervals) and MAP estimates. For comparison, literature estimates and confidence intervals are also included, where permitted by the abscissa range.

			Linear and nonlinear dependencies between a pair of parameters may be qualitatively judged by means of the joint marginal posteriors (\secref{sec_posterior_inference}) shown in \figref{fig_res_JMP}. The inclinations of their equiprobability contours with respect to the coordinate axes are related to the corresponding correlation coefficients.
			Some joint marginal posteriors exhibit clear deviations from bivariate normal distributions, illustrating that a Gaussian approximation of the likelihood by use of the Fischer matrix (\secref{sec_freq_inference}) would be inappropriate.
			
			\Tabref{tab_res_corr_coeff} lists the final posterior correlation coefficients. When these attain high absolute values, model-related equations can sometimes serve as an explanation.
			
			For example, the highly negative correlation between $\PMf$ and $\PMchi$ is related to \eqref{eq_Keplers_eq_chi} in the following way. Given data and estimated model parameters, let $\tm_{\mathrm m}$ be the time midway between the first and last observations, and $\EA_{\mathrm m}$ the eccentric anomaly at $\tm_{\mathrm m}$. Now increase $\PMf$ by a small amount. This can be balanced by a small decrease of $\PMchi$ (or, equivalently, by increasing the time of periapsis~$\PMT$) such that the eccentric anomaly at $\tm_{\mathrm m}$ again equals $\EA_{\mathrm m}$. This contrarious change of $\PMf$ and $\PMchi$, as opposed to leaving only one of them altered, will make $\EA$ deviate less, on average, over the total timespan; \ie\ the model will fit the data better. The relation between $\PMf$ and $\PMchi$ so introduced is indicated by their negative correlation coefficient.
			
			As another example, the highly negative correlation coefficient $\PMomegaii$ and $\PMOmega$ can be understood by reference to \eqsrefrange{eq_rot_matrix_A}{eq_rot_matrix_G}, which describes the Thiele-Innes constants of the AM model as follows. In the case of edge-on orbits (inclination $\PMi = 0$), these expressions simplify to sums or differences of $\cos(\cdot)$ and $\sin(\cdot)$ functions, the arguments being $\PMomegaii+\PMOmega$ or $\PMomegaii-\PMOmega$, with both arguments appearing equally often. An additional simplification is observed for face-on orbits ($\PMi = \frac{\pi}{2}$), where the Thiele-Innes constants are $A = G = -\cos(\PMOmega+\PMomegaii)$ and $B = -F = -\sin(\PMOmega+\PMomegaii)$. Thus, for orbits nearly face-on, the strong appearance of the sum $\PMOmega+\PMomegaii$ introduces a negative correlation between the two angles, because their contrarious change can lead to the same value of the model function. For Doppler-spectroscopic data, this is counteracted by the fact that the RV model function does not contain $\PMOmega$.

			\paragraph{Models and residuals.}
			\Figref{fig_res_da_mo_re_all} presents all data with the models calculated from the MAP estimates listed in \tabref{tab_res_pars}, as well as residual vectors and error ellipses for astrometry. While this analysis used both the older \citep{280} and newer \citep{260} AM in combination with RV data \citep{290}, the AM model is very similar to the one calculated by \citet[][\refWordFig~11]{260} using only the newer AM data because of the overall agreement in parameters.
			
			\Figref{fig_res_da_mo_re_all_fold} shows all data, models and residuals, separately by coordinate for AM. The abscissa corresponds to the mean anomaly~$\MA$ (\eqref{eq_Keplers_eq_chi}), \ie\ the plots are \term{folded} \wrt\ a time of periapsis~$\PMT$ and period~$\PMP$ corresponding to the posterior mode. Again, due to the similarity in parameters, the folded RV plot in \figref{fig_res_da_mo_re_all_fold} is very similar to the corresponding figure in \citet{290}.
			
			\begin{table}
				\caption{$p$-values of the Kolmogorov-Smirnov statistic.\tfm{a}}\lbl{tab_res_p_vals}
				\centering
				\begin{tabular}{lr@{.}l}
					\hline\hline
					Data type & \multicolumn{2}{l}{$p$-value}\\\hline
					AM & 0 & 00098\\
					RV & 0 & 75139\\
					All data & 0 & 00331\\
					\hline
				\end{tabular}
				\tablefoot{
					\tablefoottext{a}{These values are defined with respect to the final residuals and a standard normal distribution.}
				}
			\end{table}
			To assess the distribution of the residuals and compare it to a normal distribution, thus checking the validity of our noise model, we normalised the residuals of both data types as follows. For AM, we defined the normalised residual as a signed version of the Mahalanobis distance \citep{671} between the observed and the modelled values,
			\begin{equation}
				\RESIDnorm_{\mathrm{AM,i}} \defby s_i\sqrt{(\vr_i-\mdlFctAM{\vPar}{\tm_i})\transp\,\dataCovM_i\inv\,(\vr_i-\mdlFctAM{\vPar}{\tm_i})},\lbl{eq_def_norm_resid_AM}
			\end{equation}
			with sign
			\begin{equation}
				s_i \defby \sgn\left((\UNCERTELLphi_i-\RESIDphi_i)(\pi-[\UNCERTELLphi_i-\RESIDphi_i])\right),
			\end{equation}
			where $\RESIDphi_i$ and $\UNCERTELLphi_i$ are the position angles of the residual and of the uncertainty ellipse (see \secref{sec_freq_inference}), respectively. This definition allows us to write, according to \eqref{eq_chi_sq_AM},
			\begin{equation}
				\chi^2\iAM = \sum_{i=1}^{N\iAM} \RESIDnorm_{\mathrm{AM,i}}^2
			\end{equation}
			For RV data, we define
			\begin{equation}
				\RESIDnorm_{\mathrm{RV,i}} \defby \frac{\RV_i-\mdlFctRV{\vPar}{\tm_i}}{\measUncToti{i}},
			\end{equation}
			which analogously yields
			\begin{equation}
				\chi^2\iRV = \sum_{i=1}^{N\iRV} \RESIDnorm_{\mathrm{RV,i}}^2.\lbl{eq_chi_sq_RV_simple}
			\end{equation}
			The distributions of normalised residuals of both data types individually as well as of all data, estimated using kernel density estimation (\secref{sec_comput_techniques}), are shown in \figref{fig_res_distr_norm_resid_all}. \Tabref{tab_res_p_vals} lists the $p$-values of the \term{Kolmogorov-Smirnov statistic}, which relates each distribution to the standard normal distribution~$\stdNormDistr$.
			
			The $p$-value equals the probability, under the hypothesis~$\hy_{\mathcal N}$ that the normalised residuals are randomly drawn from $\stdNormDistr$, to observe a distribution of normalised residuals that differs at least as much from $\stdNormDistr$ as is actually the case. This difference is quantified here by the Kolmogorov-Smirnov statistic. Denoting the hypothesis of such an observation by $\hy_{\mathrm{obs}}$, the $p$-value can be expressed as $\p(\hy_{\mathrm{obs}}|\hy_{\mathcal N})$. We note that according to the Bayes theorem, this is not equal to the ``inverse'' probability~$\p(\hy_{\mathcal N}|\hy_{\mathrm{obs}})$ of the residuals coming from the normal distribution, given the observation.
			
			For the AM residuals, as well as for all residuals, a heavy-tailed distribution (\figref{fig_res_distr_norm_resid_all}) is observed and the low $p$-value indicates a minute probability of such an observation under $\hy_{\mathcal N}$, \ie\ the normalised residuals comply poorly with the standard normal distribution~$\stdNormDistr$. This reflects the fact that several AM data points are outliers \wrt\ the observable and noise model (\figref{fig_res_da_mo_re_all}). We interpret this in terms of systematic effects in the measurements that are not contained in our noise model. In contrast, the normalised RV residuals are more normally-distributed, and there are no such severe outliers (\figref{fig_res_da_mo_re_all} and \figref{fig_res_da_mo_re_all_fold}).
			
			According to the principle of maximum entropy, given only the mean and variance of a distribution, the normal distribution has maximum information-theoretic entropy, equivalent to minimum bias or prejudice with respect to the missing information \citepeg{672}. Still, it is well-known that this widely used noise model is relatively prone to outliers; \citet{670} have suggested to replace it by a non-standardised $t$-distribution, resulting in the down-weighting of outliers. This distribution can be derived from the normal distribution under the assumption that the noise variance is unknown with a certain probability distribution. The $t$-distribution has an additional unknown degrees-of-freedom parameter~$\nu\in\Real$; for $\nu=1$, it resembles the Cauchy-distribution.

			In contrast, our approach has been to regard every datum with its standard deviation as accurate, which also implies that we have not discarded any data as outliers. Under this assumption, deviating from the maximum-entropy principle by selecting a different distribution introduces prior ``knowledge'' that we may not actually have and thus potentially biases the results.

%
	\section{Conclusions}\lbl{sec_conclus}
%

	We have presented BASE, a novel and highly configurable tool for Bayesian parameter and uncertainty estimation with respect to model parameters and additional derived quantities, which can be applied to AM as well as RV data of both exoplanet systems and binary stars. With user-specified or uninformative prior knowledge, it employs a combination of \MCMClong\ (MCMC) and several other techniques to explore the whole parameter space and collect samples distributed according to the posterior distribution. We presented a new, simple method of refining the window width of one-dimensional kernel density estimation, which is used to derive marginal posterior densities.
	
	We derived the observable models from Newton's law of gravitation, neglecting the motion of the observer and the target system, which we showed is justified in the case of \miz. After sketching how we estimated the RV uncertainties that are missing in the original publication \citep{290}, we detailed our analysis of all publicly available AM and RV data of \miz. It consists of five consecutive stages and has produced estimates of the values, uncertainties, and correlations of all model parameters and derived quantities, as well as marginal posterior densities over one and two dimensions.

	As illustrated in \figref{fig_res_mapo} and \tabref{tab_res_pars}, our new results exhibit overall compatibility with previous literature values; this is also the case for the models in \figref{fig_res_da_mo_re_all}, whose plots differ only slightly from those published earlier. Several outliers in the AM data are visible in the distribution of the corresponding normalised residuals, which deviates significantly more from a standard normal distribution than that of the RV residuals. Nevertheless, it is not necessary to remove outliers for our program to finish successfully.
	
	In the near future, we plan to apply BASE to a potential exoplanet host star. In this study one of the aims will be to determine the existence probability of a planetary companion.

%
%
																															\appendix
%
%

%
	\section{Encoding prior knowledge}\lbl{sec_priors}
%

		By means of the prior, Bayesian analysis allows one to incorporate knowledge obtained earlier, \eg\ using different data. When no prior knowledge is available for some model parameter, except for its allowed range, maximum prior ignorance about the parameter can be encoded by a prior of one of the following functional forms for the most common classes of \term{location} and \term{scale parameters} \citep{075,102}.
		\begin{itemize}
			\item For a \term{location parameter}, we demand that the prior be invariant against a shift~$\Delta$ in the parameter, \ie
			\begin{equation}
				\pPri{\param}\ddd\param = \pPri{\param+\Delta}\ddd(\param+\Delta),
			\end{equation} 
			which leads to the \term{uniform} prior
			\begin{equation}
				\pPri{\param} = \uniformPrior{\param}{a}{b},
			\end{equation}
			where $\Theta(\cdot), a,$ and $b$ are the Heaviside step function, the lower and the upper prior bounds.
			
			Here, we note that the frequentist approach, lacking an explicit definition of the prior, corresponds to the implicit assumption of a uniform prior for all parameters.
			
			\item A positive \term{scale parameter}, which often spans several decades, is characterised by its invariance against a stretch of the coordinate axis by a factor~$\varphi$, \ie
			\begin{equation}
				\pPri{\param}\ddd\param = \pPri{\varphi\param}\ddd(\varphi\param),
			\end{equation}
			which is solved by the \term{Jeffreys prior},
			\begin{equation}
				\pPri{\param} = \jeffreysPrior{\param}{a}{b}.
			\end{equation}
			That a uniform prior would be inappropriate for this parameter is also illustrated by the fact that it would assign higher probabilities to $\param$ lying in a higher decade of $[a,b]$ than in a lower.
			
			\item If the lower prior bound of a scale parameter is zero, \eg\ for the RV semi-amplitude~$\PMK$, a \term{modified Jeffreys prior} is used. It has the form
			\begin{equation}
				\pPri{\param} = \modJeffreysPrior{\param}{\knee{\param}}{a}{b},
			\end{equation}
			where $\knee{\param}$ is the \term{knee} of the prior. For $\param\ll\knee{\param}$, this prior is approximately uniform, while it approaches a Jeffreys prior for $\param\gg\knee{\param}$.
		\end{itemize}

%
	\section{Numerical posterior summaries}\lbl{sec_num_summaries}
%

		The following tables list the numerical values of several posterior summaries. Those in \tabsrefii{tab_res_pars}{tab_res_DQs} are derived from the marginal posteriors (\figref{fig_res_mapo}), while the correlation coefficients in \tabref{tab_res_corr_coeff} reflect linear relations between parameters and, in this respect, can be regarded as summaries of the joint marginal posteriors (\figref{fig_res_JMP}).
		
		Compared to the underlying densities, all of these summaries are incomplete. Still, they are useful \eg\ for the calculation of model functions or comparison with literature results.

				
				\begin{table*}
					\caption{Model parameters: new and previous results.\tfm{a}}
					\lbl{tab_res_pars}
					\centering
					\begin{tabular}{lr@{.}lr@{.}lr@{.}lr@{.}lr@{.}lr@{.}lr@{.}lr@{.}lr@{.}lr@{.}ll}\hline\hline
						Estimate &\multicolumn{2}{l}{$\PMe$} & \multicolumn{2}{l}{$\PMf$} & \multicolumn{2}{l}{$\PMchi$} & \multicolumn{2}{l}{$\PMomegaii$} & \multicolumn{2}{l}{$\PMi$} & \multicolumn{2}{l}{$\PMOmega$} & \multicolumn{2}{l}{$\PMpi$} & \multicolumn{2}{l}{$\PMV$} & \multicolumn{2}{l}{$\PMKi$} & \multicolumn{2}{l}{$\PMKii$} & Reference\\
						& \multicolumn{2}{l}{} & \multicolumn{2}{l}{$(\UNITdInvFlat)$} & \multicolumn{2}{l}{} & \multicolumn{2}{l}{$(\UNITrad)$} & \multicolumn{2}{l}{$(\UNITrad)$} & \multicolumn{2}{l}{$(\UNITrad)$} & \multicolumn{2}{l}{$(\UNITmas)$} & \multicolumn{2}{l}{$(\UNITkmpsFlat)$} & \multicolumn{2}{l}{$(\UNITkmpsFlat)$} & \multicolumn{2}{l}{$(\UNITkmpsFlat)$}\\\hline
						$\md\param$     & $0$ & $5304$ & $0$ & $0486\,89388$ & $0$ & $93322$ & $4$ & $9771$ & $1$ & $0530$ & $1$ & $8381$ & $38$ & $74$ &  $-6$ & $02$ & $58$ & $84$ & $57$ & $16$\\
						$\margMd\param$ & $0$ & $5299$ & $0$ & $0486\,89403$ & $0$ & $93315$ & $4$ & $9779$ & $1$ & $0528$ & $1$ & $8374$ & $38$ & $69$ &  $-6$ & $20$ & $58$ & $60$ & $57$ & $39$\\
						$\med\param$     &$0$ & $5295$ & $0$ & $0486\,89409$ & $0$ & $93294$ & $4$ & $9784$ & $1$ & $0527$ & $1$ & $8365$ & $38$ & $91$ &  $-6$ & $04$ & $58$ & $41$ & $56$ & $97$\\
						$\mn\param$     & $0$ & $5281$ & $0$ & $0486\,89438$ & $0$ & $93244$ & $4$ & $9807$ & $1$ & $0515$ & $1$ & $8346$ & $39$ & $02$ &  $-6$ & $02$ & $58$ & $21$ & $56$ & $83$\\

						\multirow{2}{*}{$\HPDIArg{50\%}$} & $0$ & $5282$ & $0$ & $0486\,89341$ & $0$ & $93227$ & $4$ & $9751$ & $1$ & $0513$ & $1$ & $8345$ & $38$ & $13$ &  $-6$ & $93$ & $57$ & $14$ & $55$ & $47$\\
						& $0$ & $5317$ & $0$ & $0486\,89467$ & $0$ & $93379$ & $4$ & $9806$ & $1$ & $0549$ & $1$ & $8397$ & $39$ & $66$ &  $-5$ & $21$ & $60$ & $21$ & $58$ & $47$\\
						\multirow{2}{*}{$\HPDIArg{68.27\%}$} & $0$ & $5270$ & $0$ & $0486\,89295$ & $0$ & $93175$ & $4$ & $9735$ & $1$ & $0500$ & $1$ & $8327$ & $37$ & $74$ &  $-7$ & $31$ & $56$ & $18$ & $54$ & $69$\\
						& $0$ & $5326$ & $0$ & $0486\,89509$ & $0$ & $93430$ & $4$ & $9825$ & $1$ & $0558$ & $1$ & $8411$ & $40$ & $06$ &  $-4$ & $68$ & $60$ & $84$ & $59$ & $23$\\
						\multirow{2}{*}{$\HPDIArg{95\%}$} & $0$ & $5152$ & $0$ & $0486\,88907$ & $0$ & $92550$ & $4$ & $9653$ & $1$ & $0383$ & $1$ & $8165$ & $35$ & $73$ &  $-9$ & $96$ & $51$ & $77$ & $50$ & $35$\\
						& $0$ & $5380$ & $0$ & $0486\,90031$ & $0$ & $93860$ & $5$ & $0019$ & $1$ & $0615$ & $1$ & $8485$ & $42$ & $24$ &  $-2$ & $08$ & $64$ & $54$ & $63$ & $16$\\
						\multirow{2}{*}{$\HPDIArg{99\%}$} & $0$ & $4905$ & $0$ & $0486\,88483$ & $0$ & $91472$ & $4$ & $9530$ & $1$ & $0134$ & $1$ & $7790$ & $34$ & $24$ & $-12$ & $40$ & $46$ & $87$ & $45$ & $66$\\
						& $0$ & $5451$ & $0$ & $0486\,90874$ & $0$ & $94285$ & $5$ & $0423$ & $1$ & $0699$ & $1$ & $8593$ & $45$ & $02$ &   $0$ & $22$ & $67$ & $05$ & $66$ & $49$\\
						Lit. estimate & $0$ & $537$ & $0$ & $0486\,8881$\,\tfm{b} & $0$ & $93487$\,\tfm{b} & $4$ & $9595$ & \multicolumn{2}{c}{\ldots} & \multicolumn{2}{c}{\ldots} & \multicolumn{2}{c}{\ldots} & $-5$ & $64$ & $58$ & $04$\,\tfm{d} & $57$ & $03$\,\tfm{d} & \multirow{2}{*}{(1)} \\
						Uncertainty & $0$ & $004$ & \multicolumn{2}{c}{\ldots\,\tfm{c}} & \multicolumn{2}{c}{\ldots\,\tfm{c}} & $0$ & $0201$ & \multicolumn{2}{c}{\ldots} & \multicolumn{2}{c}{\ldots} & \multicolumn{2}{c}{\ldots} & $0$ & $15$ & $0$ & $70$ & $0$ & $80$\\
						Lit. estimate & $0$ & $5354$ & $0$ & $0486\,89403$\,\tfm{b} & $0$ & $93524$\,\tfm{b} & $4$ & $9620$ & $1$ & $0559$ & $1$ & $850$ & $39$ & $4$ & \multicolumn{2}{c}{\ldots} & $58$ & $33$\,\tfm{e} & $56$ & $69$\,\tfm{e} & \multirow{2}{*}{(2)} \\
						Uncertainty & $0$ & $0025$ & $0$ & $0000\,00119$ & $0$ & $00097$ & $0$ & $0052$ & $0$ & $0052$ & $0$ & $007$ & $0$ & $3$ & \multicolumn{2}{c}{\ldots} & $2$ & $40$ & $2$ & $33$\\
						\hline
					\end{tabular}
					\tablefoot{
						\tablefoottext{a}{For definitions of the estimates, see \secref{sec_posterior_inference}.}
						\tablefoottext{b}{Literature values and uncertainties are calculated from $\tmOne$ and the original parameters $\PMP$ and $\PMT$ according to \tabref{tab_DQs}.}
						\tablefoottext{c}{Uncertainties are missing for $\PMP$ in \citet{290} and thus for $\PMf$ and $\PMchi$.}
						\tablefoottext{d}{$\PMKi$ and $\PMKii$ are derived from $\PMKLiti{1}$ and $\PMKLiti{2}$ via $\PMe$ according to \tabref{tab_DQs}.}
						\tablefoottext{e}{$\PMKi$ and $\PMKii$ are derived from $\PMP,\PMaRelApp,\PMpi,\PMrho$ and $\PMi$ using \eqsrefiii{eq_def_CM}{eq_RV_amplitude_II}{eq_def_a_rel_app}.}
					}
					\tablebib{(1)~\citet{290}; (2)~\citet{260}}
				\end{table*}

				\begin{table*}
					\caption{Derived quantities: new and previous results.\tfm{a}}
					\lbl{tab_res_DQs}
					\centering
					\begin{tabular}{lr@{.}lr@{.}lr@{.}lr@{.}lr@{.}lr@{.}lr@{.}lr@{.}lr@{.}lr@{.}lr@{.}ll}
						\hline\hline
						Estimate & \multicolumn{2}{l}{$\PMP$} & \multicolumn{2}{l}{$\PMT$\,\tfm{b}} & \multicolumn{2}{l}{$\PMd$} & \multicolumn{2}{l}{$\PMrho$} & \multicolumn{2}{l}{$\PMKLiti{1}$} & \multicolumn{2}{l}{$\PMKLiti{2}$} & \multicolumn{2}{l}{$\PMmC{1}$} & \multicolumn{2}{l}{$\PMmC{2}$} & \multicolumn{2}{l}{$\PMaC{1}$} & \multicolumn{2}{l}{$\PMaC{2}$} & \multicolumn{2}{l}{$\PMaRel$} & Reference\\
						& \multicolumn{2}{l}{$(\UNITd)$} & \multicolumn{2}{l}{$(\UNITd)$} & \multicolumn{2}{l}{$(\UNITpc)$} & \multicolumn{2}{l}{} & \multicolumn{2}{l}{$(\UNITkmpsFlat)$} & \multicolumn{2}{l}{$(\UNITkmpsFlat)$} & \multicolumn{2}{l}{$(\UNITMSun)$} & \multicolumn{2}{l}{$(\UNITMSun)$} & \multicolumn{2}{l}{$(\UNITAU)$} & \multicolumn{2}{l}{$(\UNITAU)$} & \multicolumn{2}{l}{$(\UNITAU)$} &\\\hline
						$\margMd\param$  & $20$ & $53\,8350$ & $36997$ & $247$ & $25$ & $74$ & $1$ & $028$ & $68$ & $70$ & $68$ & $05$ & $2$ & $477$ & $2$ & $500$ & $0$ & $12657$ & $0$ & $12418$ & $0$ & $25074$\\
						$\med\param$  & $20$ & $53\,8347$ & $36997$ & $251$ & $25$ & $69$ & $1$ & $025$ & $68$ & $85$ & $67$ & $15$ & $2$ & $459$ & $2$ & $517$ & $0$ & $12678$ & $0$ & $12353$ & $0$ & $25068$\\
						$\mn\param$   & $20$ & $53\,8335$ & $36997$ & $262$ & $25$ & $66$ & $1$ & $027$ & $68$ & $56$ & $66$ & $93$ & $2$ & $455$ & $2$ & $521$ & $0$ & $12658$ & $0$ & $12393$ & $0$ & $25017$\\

						\multirow{2}{*}{$\HPDIArg{50\%}$} & $20$ & $53\,8322$ & $36997$ & $234$ & $25$ & $18$ & $0$ & $986$ & $67$ & $30$ & $65$ & $39$ & $2$ & $320$ & $2$ & $311$ & $0$ & $12331$ & $0$ & $11742$ & $0$ & $24628$\\
						& $20$ & $53\,8375$ & $36997$ & $265$ & $26$ & $20$ & $1$ & $061$ & $70$ & $90$ & $68$ & $91$ & $2$ & $609$ & $2$ & $689$ & $0$ & $13051$ & $0$ & $12987$ & $0$ & $25555$\\
						\multirow{2}{*}{$\HPDIArg{68.27\%}$} & $20$ & $53\,8305$ & $36997$ & $223$ & $24$ & $94$ & $0$ & $969$ & $66$ & $26$ & $64$ & $48$ & $2$ & $238$ & $2$ & $228$ & $0$ & $12156$ & $0$ & $11385$ & $0$ & $24388$\\
						& $20$ & $53\,8395$ & $36997$ & $276$ & $26$ & $47$ & $1$ & $082$ & $71$ & $71$ & $69$ & $82$ & $2$ & $678$ & $2$ & $809$ & $0$ & $13269$ & $0$ & $13293$ & $0$ & $25803$\\
						\multirow{2}{*}{$\HPDIArg{95\%}$} & $20$ & $53\,8085$ & $36997$ & $135$ & $23$ & $46$ & $0$ & $866$ & $60$ & $95$ & $59$ & $10$ & $1$ & $827$ & $1$ & $775$ & $0$ & $11252$ & $0$ & $10058$ & $0$ & $22977$\\
						& $20$ & $53\,8558$ & $36997$ & $404$ & $27$ & $72$ & $1$ & $187$ & $75$ & $99$ & $74$ & $19$ & $3$ & $051$ & $3$ & $235$ & $0$ & $14028$ & $0$ & $14681$ & $0$ & $26908$\\
						\multirow{2}{*}{$\HPDIArg{99\%}$} & $20$ & $53\,7729$ & $36997$ & $044$ & $21$ & $98$ & $0$ & $751$ & $54$ & $61$ & $53$ & $79$ & $1$ & $475$ & $1$ & $463$ & $0$ & $10452$ & $0$ & $08958$ & $0$ & $21512$\\
						& $20$ & $53\,8737$ & $36997$ & $622$ & $28$ & $92$ & $1$ & $305$ & $78$ & $41$ & $78$ & $28$ & $3$ & $415$ & $3$ & $641$ & $0$ & $14637$ & $0$ & $16321$ & $0$ & $27939$\\
						Lit. estimate & $20$ & $53\,860$ & $36997$ & $212$ & \multicolumn{2}{c}{\ldots} & $1$ & $018$ & $68$ & $80$ & $67$ & $60$ & \multicolumn{2}{c}{\ldots} & \multicolumn{2}{c}{\ldots} & \multicolumn{2}{c}{\ldots} & \multicolumn{2}{c}{\ldots} & \multicolumn{2}{c}{\ldots} & \multirow{2}{*}{(1)}\\
						Uncertainty & \multicolumn{2}{c}{\ldots} & $0$ & $022$ & \multicolumn{2}{c}{\ldots} & $0$ & $018$ & $0$ & $79$ & $0$ & $91$ & \multicolumn{2}{c}{\ldots} & \multicolumn{2}{c}{\ldots} & \multicolumn{2}{c}{\ldots} & \multicolumn{2}{c}{\ldots} & \multicolumn{2}{c}{\ldots} & \\
						Lit. estimate & $20$ & $53\,835$ & $36997$ & $20$ & $25$ & $38$ & $1$ & $029$ & $69$ & $06$ & $67$ & $13$ & $2$ & $43$ & $2$ & $50$ & $0$ & $12652$ & $0$ & $12297$ & $0$ & $24949$ & \multirow{2}{*}{(2)}\\
						Uncertainty & $0$ & $00\,005$ & $0$ & $03$ & $0$ & $19$ & $0$ & $041$ & $3$ & $84$ & $3$ & $74$ & $0$ & $07$ & $0$ & $07$ & $0$ & $00519$ & $0$ & $00504$ & $0$ & $00205$ & \\
						\hline
					\end{tabular}
					\tablefoot{
						\tablefoottext{a}{For definitions of the estimates, see \secref{sec_posterior_inference}.}
						\tablefoottext{b}{$\PMT$ is given in the \term{reduced Julian date} scale, \ie\ as Julian Date$\,-\,\EE{2.4}{6}\UNITd$.}
					}
					\tablebib{(1)~\citet{290}; (2)~\citet{260}}
				\end{table*}

				\begin{table*}
					\caption{Pearson correlation coefficients of pairs of parameters.}\lbl{tab_res_corr_coeff}
					\centering
					\begin{tabular}{l||r@{.}lr@{.}lr@{.}lr@{.}lr@{.}lr@{.}lr@{.}lr@{.}lr@{.}lr@{.}l}
						\hline\hline
						& \multicolumn{2}{l}{$\PMe$} & \multicolumn{2}{l}{$\PMf$} & \multicolumn{2}{l}{$\PMchi$} & \multicolumn{2}{l}{$\PMomegaii$} & \multicolumn{2}{l}{$\PMi$} & \multicolumn{2}{l}{$\PMOmega$} & \multicolumn{2}{l}{$\PMpi$} & \multicolumn{2}{l}{$\PMV$} & \multicolumn{2}{l}{$\PMKi$}\\\hline
							$\PMf$       & $-0$ & $21017$ & \multicolumn{2}{c}{\ldots} & \multicolumn{2}{c}{\ldots} & \multicolumn{2}{c}{\ldots} & \multicolumn{2}{c}{\ldots} & \multicolumn{2}{c}{\ldots} & \multicolumn{2}{c}{\ldots} & \multicolumn{2}{c}{\ldots} & \multicolumn{2}{c}{\ldots}\\
							$\PMchi$     &  $0$ & $35120$ & $-0$ & $43922$ & \multicolumn{2}{c}{\ldots} & \multicolumn{2}{c}{\ldots} & \multicolumn{2}{c}{\ldots} & \multicolumn{2}{c}{\ldots} & \multicolumn{2}{c}{\ldots} & \multicolumn{2}{c}{\ldots} & \multicolumn{2}{c}{\ldots}\\
							$\PMomegaii$ & $-0$ & $62560$ &  $0$ & $18818$ & $-0$ & $35074$ & \multicolumn{2}{c}{\ldots} & \multicolumn{2}{c}{\ldots} & \multicolumn{2}{c}{\ldots} & \multicolumn{2}{c}{\ldots} & \multicolumn{2}{c}{\ldots} & \multicolumn{2}{c}{\ldots}\\
							$\PMi$       &  $0$ & $52716$ & $-0$ & $11723$ &  $0$ & $28022$ & $-0$ & $57440$ & \multicolumn{2}{c}{\ldots} & \multicolumn{2}{c}{\ldots} & \multicolumn{2}{c}{\ldots} & \multicolumn{2}{c}{\ldots} & \multicolumn{2}{c}{\ldots}\\
							$\PMOmega$   &  $0$ & $60484$ & $-0$ & $22616$ &  $0$ & $34159$ & $-0$ & $62355$ &  $0$ & $47121$ & \multicolumn{2}{c}{\ldots} & \multicolumn{2}{c}{\ldots} & \multicolumn{2}{c}{\ldots} & \multicolumn{2}{c}{\ldots}\\
							$\PMpi$      &  $0$ & $02224$ &  $0$ & $04643$ & $-0$ & $03317$ & $-0$ & $02066$ &  $0$ & $05501$ & $-0$ & $00134$ & \multicolumn{2}{c}{\ldots} & \multicolumn{2}{c}{\ldots} & \multicolumn{2}{c}{\ldots}\\
							$\PMV$       & $-0$ & $01699$ &  $0$ & $02057$ & $-0$ & $02176$ &  $0$ & $01468$ & $-0$ & $01012$ & $-0$ & $01436$ &  $0$ & $00752$ & \multicolumn{2}{c}{\ldots} & \multicolumn{2}{c}{\ldots}\\
							$\PMKi$      &  $0$ & $11517$ & $-0$ & $07289$ &  $0$ & $09702$ & $-0$ & $12194$ &  $0$ & $09822$ &  $0$ & $11416$ & $-0$ & $33947$ &  $0$ & $00868$ & \multicolumn{2}{c}{\ldots}\\
							$\PMKii$     &  $0$ & $04381$ & $-0$ & $03506$ &  $0$ & $04188$ & $-0$ & $04901$ &  $0$ & $03520$ &  $0$ & $04618$ & $-0$ & $32930$ & $-0$ & $02639$ & $0$ & $02274$\\
						\hline
					\end{tabular}
				\end{table*}

	\begin{acknowledgements}
		We wish to thank David~W. Hogg, Sabine Reffert, Ren\'e Andrae, and Mathias Zechmeister for fruitful discussions, and an anonymous referee for comments that have improved the quality and clarity of this article.
		This research has made use of the SIMBAD database and VizieR catalogue access tool, operated at CDS, Strasbourg, France, as well as NASA's Astrophysics Data System.
	\end{acknowledgements}

	\bibliography{references_aa}

\begin{thebibliography}{68}
\expandafter\ifx\csname natexlab\endcsname\relax\def\natexlab#1{#1}\fi

\bibitem[{{Adamson} {et~al.}(2005){Adamson}, {Aspin}, {Davis}, {Fujiyoshi}, \&
  {A.~Adamson, C.~Aspin, C.~Davis, \& T.~Fujiyoshi}}]{488}
{Adamson}, A., {Aspin}, C., {Davis}, C., {Fujiyoshi}, T., \& {A.~Adamson,
  C.~Aspin, C.~Davis, \& T.~Fujiyoshi}, eds. 2005, Astronomical Society of the
  Pacific Conference Series, Vol. 343, {Astronomical Polarimetry: Current
  Status and Future Directions}

\bibitem[{{Armstrong} {et~al.}(1998){Armstrong}, {Mozurkewich}, {Rickard},
  {Hutter}, {Benson}, {Bowers}, {Elias}, {Hummel}, {Johnston}, {Buscher},
  {Clark}, {Ha}, {Ling}, {White}, \& {Simon}}]{297}
{Armstrong}, J.~T., {Mozurkewich}, D., {Rickard}, L.~J., {et~al.} 1998, \apj,
  496, 550

\bibitem[{{Armstrong} {et~al.}(1992){Armstrong}, {Mozurkewich}, {Vivekanand},
  {Simon}, {Denison}, {Johnston}, {Pan}, {Shao}, \& {Colavita}}]{284}
{Armstrong}, J.~T., {Mozurkewich}, D., {Vivekanand}, M., {et~al.} 1992, \aj,
  104, 241

\bibitem[{Bayes \& Price(1763)}]{464}
Bayes, M. \& Price, M. 1763, Philosophical Transactions, 53, 370

\bibitem[{{Bond}(1857)}]{460}
{Bond}, G. 1857, \mnras, 17, 230

\bibitem[{Boneh \& Golan(1979)}]{599}
Boneh, A. \& Golan, A. 1979, in Third European Congress on Operations Research,
  EURO III, Amsterdam

\bibitem[{Brooks \& Gelman(1998)}]{322}
Brooks, S.~P. \& Gelman, A. 1998, Journal of Computational and Graphical
  Statistics, 7, 434

\bibitem[{{Charbonneau} {et~al.}(2000){Charbonneau}, {Brown}, {Latham}, \&
  {Mayor}}]{484}
{Charbonneau}, D., {Brown}, T.~M., {Latham}, D.~W., \& {Mayor}, M. 2000, \apjl,
  529, L45

\bibitem[{Deeg {et~al.}(2007)Deeg, Belmonte, \& Aparicio}]{433}
Deeg, H.~J., Belmonte, J.~A., \& Aparicio, A., eds. 2007, Extrasolar Planets
  (Cambridge University Press)

\bibitem[{{Delplancke}(2008)}]{496}
{Delplancke}, F. 2008, \nar, 52, 199

\bibitem[{{Delplancke} {et~al.}(2000){Delplancke}, {Leveque}, {Kervella},
  {Glindemann}, \& {D'Arcio}}]{495}
{Delplancke}, F., {Leveque}, S.~A., {Kervella}, P., {Glindemann}, A., \&
  {D'Arcio}, L. 2000, in Presented at the Society of Photo-Optical
  Instrumentation Engineers (SPIE) Conference, Vol. 4006, Society of
  Photo-Optical Instrumentation Engineers (SPIE) ConferenceSeries, ed.
  {P.~L{\'e}na \& A.~Quirrenbach}, 365--376

\bibitem[{Efron \& Tibshirani(1993)}]{524}
Efron, B. \& Tibshirani, R. 1993, An introduction to the bootstrap, Monographs
  on statistics and applied probability (Chapman \& Hall)

\bibitem[{{Ford}(2004)}]{096}
{Ford}, E.~B. 2004, in American Institute of Physics Conference Series, Vol.
  713, The Search for Other Worlds, ed. S.~S. {Holt} \& D.~{Deming}, 27--30

\bibitem[{{Ford}(2006)}]{023}
{Ford}, E.~B. 2006, \apj, 642, 505

\bibitem[{{Free Software Foundation}(2011{\natexlab{a}})}]{513}
{Free Software Foundation}. 2011{\natexlab{a}}, {GFortran},
  http://gcc.gnu.org/fortran/

\bibitem[{{Free Software Foundation}(2011{\natexlab{b}})}]{512}
{Free Software Foundation}. 2011{\natexlab{b}}, {GOMP},
  http://gcc.gnu.org/projects/gomp/

\bibitem[{{Gatewood} {et~al.}(1980){Gatewood}, {Breakiron}, {Goebel}, {Kipp},
  {Russell}, \& {Stein}}]{492}
{Gatewood}, G., {Breakiron}, L., {Goebel}, R., {et~al.} 1980, \icarus, 41, 205

\bibitem[{Gelman \& Rubin(1992)}]{514}
Gelman, A. \& Rubin, D.~B. 1992, Statistical Science, 7, pp. 457

\bibitem[{Geman \& Geman(1984)}]{552}
Geman, S. \& Geman, D. 1984, Pattern Analysis and Machine Intelligence, IEEE
  Transactions on, PAMI-6, 721

\bibitem[{{Gilks} {et~al.}(1996){Gilks}, {Richardson}, \&
  {Spiegelhalter}}]{098}
{Gilks}, W.~R., {Richardson}, S., \& {Spiegelhalter}, D.~J. 1996, Markov Chain
  Monte Carlo in Practice, 1st edn. (London: Chapman \& Hall)

\bibitem[{{Gillessen} {et~al.}(2010){Gillessen}, {Eisenhauer}, {Perrin},
  {Brandner}, {Straubmeier}, {Perraut}, {Amorim}, {Sch{\"o}ller},
  {Araujo-Hauck}, {Bartko}, {Baumeister}, {Berger}, {Carvas}, {Cassaing},
  {Chapron}, {Choquet}, {Clenet}, {Collin}, {Eckart}, {Fedou}, {Fischer},
  {Gendron}, {Genzel}, {Gitton}, {Gonte}, {Gr{\"a}ter}, {Haguenauer}, {Haug},
  {Haubois}, {Henning}, {Hippler}, {Hofmann}, {Jocou}, {Kellner}, {Kervella},
  {Klein}, {Kudryavtseva}, {Lacour}, {Lapeyrere}, {Laun}, {Lena}, {Lenzen},
  {Lima}, {Moratschke}, {Moch}, {Moulin}, {Naranjo}, {Neumann}, {Nolot},
  {Paumard}, {Pfuhl}, {Rabien}, {Ramos}, {Rees}, {Rohloff}, {Rouan}, {Rousset},
  {Sevin}, {Thiel}, {Wagner}, {Wiest}, {Yazici}, \& {Ziegler}}]{424}
{Gillessen}, S., {Eisenhauer}, F., {Perrin}, G., {et~al.} 2010, in Presented at
  the Society of Photo-Optical Instrumentation Engineers (SPIE) Conference,
  Vol. 7734, Society of Photo-Optical Instrumentation Engineers (SPIE)
  Conference Series

\bibitem[{{Gould}(2009)}]{440}
{Gould}, A. 2009, in Astronomical Society of the Pacific Conference Series,
  Vol. 403, Astronomical Society of the Pacific Conference Series, ed.
  {K.~Z.~Stanek}, 86--+

\bibitem[{{Gregory}(2005{\natexlab{a}})}]{007}
{Gregory}, P.~C. 2005{\natexlab{a}}, \apj, 631, 1198

\bibitem[{{Gregory}(2005{\natexlab{b}})}]{075}
{Gregory}, P.~C. 2005{\natexlab{b}}, {Bayesian Logical Data Analysis for the
  Physical Sciences: A Comparative Approach with `Mathematica' Support}
  (Cambridge: Cambridge University Press)

\bibitem[{{Hastings}(1970)}]{548}
{Hastings}, W.~K. 1970, Biometrika, 57, 97

\bibitem[{{Hoffleit} \& {Jaschek}(1982)}]{459}
{Hoffleit}, D. \& {Jaschek}, C. 1982, {The Bright Star Catalogue} (Yale
  University Observatory)

\bibitem[{{Holman} \& {Murray}(2005)}]{445}
{Holman}, M.~J. \& {Murray}, N.~W. 2005, Science, 307, 1288

\bibitem[{{Hummel} {et~al.}(1995){Hummel}, {Armstrong}, {Buscher},
  {Mozurkewich}, {Quirrenbach}, \& {Vivekanand}}]{280}
{Hummel}, C.~A., {Armstrong}, J.~T., {Buscher}, D.~F., {et~al.} 1995, \aj, 110,
  376

\bibitem[{{Hummel} {et~al.}(1998){Hummel}, {Mozurkewich}, {Armstrong},
  {Hajian}, {Elias}, \& {Hutter}}]{260}
{Hummel}, C.~A., {Mozurkewich}, D., {Armstrong}, J.~T., {et~al.} 1998, \aj,
  116, 2536

\bibitem[{Kapur(1989)}]{672}
Kapur, J. 1989, Maximum-Entropy Models in Science and Engineering (Wiley)

\bibitem[{{Koch} {et~al.}(2010){Koch}, {Borucki}, {Basri}, {Batalha}, {Brown},
  {Caldwell}, {Christensen-Dalsgaard}, {Cochran}, {DeVore}, {Dunham},
  {Gautier}, {Geary}, {Gilliland}, {Gould}, {Jenkins}, {Kondo}, {Latham},
  {Lissauer}, {Marcy}, {Monet}, {Sasselov}, {Boss}, {Brownlee}, {Caldwell},
  {Dupree}, {Howell}, {Kjeldsen}, {Meibom}, {Morrison}, {Owen}, {Reitsema},
  {Tarter}, {Bryson}, {Dotson}, {Gazis}, {Haas}, {Kolodziejczak}, {Rowe}, {Van
  Cleve}, {Allen}, {Chandrasekaran}, {Clarke}, {Li}, {Quintana}, {Tenenbaum},
  {Twicken}, \& {Wu}}]{641}
{Koch}, D.~G., {Borucki}, W.~J., {Basri}, G., {et~al.} 2010, \apjl, 713, L79

\bibitem[{{Kuhn} {et~al.}(2001){Kuhn}, {Potter}, \& {Parise}}]{485}
{Kuhn}, J.~R., {Potter}, D., \& {Parise}, B. 2001, \apjl, 553, L189

\bibitem[{Lange {et~al.}(1989)Lange, Little, \& Taylor}]{670}
Lange, K.~L., Little, R. J.~A., \& Taylor, J. M.~G. 1989, Journal of the
  American Statistical Association, 84, pp. 881

\bibitem[{{Launhardt} {et~al.}(2008){Launhardt}, {Queloz}, {Henning},
  {Quirrenbach}, {Delplancke}, {Andolfato}, {Baumeister}, {Bizenberger},
  {Bleuler}, {Chazelas}, {D{\'e}rie}, {Di Lieto}, {Duc}, {Duvanel}, {Elias},
  {Fluery}, {Geisler}, {Gillet}, {Graser}, {Koch}, {K{\"o}hler}, {Maire},
  {M{\'e}gevand}, {Michellod}, {Moresmau}, {M{\"u}ller}, {M{\"u}llhaupt},
  {Naranjo}, {Pepe}, {Reffert}, {Sache}, {S{\'e}gransan}, {Salvad{\'e}},
  {Schulze-Hartung}, {Setiawan}, {Simond}, {Sosnowska}, {Stilz}, {Tubbs},
  {Wagner}, {Weber}, {Weise}, \& {Zago}}]{389}
{Launhardt}, R., {Queloz}, D., {Henning}, T., {et~al.} 2008, in Society of
  Photo-Optical Instrumentation Engineers (SPIE) Conference Series, Vol. 7013,
  Society of Photo-Optical Instrumentation Engineers (SPIE) Conference Series

\bibitem[{{Levine} {et~al.}(2009){Levine}, {Soummer}, {Arenberg}, {Belikov},
  {Bierden}, {Boccaletti}, {Brown}, {Burrows}, {Burrows}, {Cady}, {Cash},
  {Clampin}, {Cossapakis}, {Crossfield}, {Dewell}, {Egerman}, {Fergusson},
  {Ge}, {Give'On}, {Guyon}, {Heap}, {Hyde}, {Jaroux}, {Jasdin}, {Kasting},
  {Kenworthy}, {Kilston}, {Klavins}, {Krist}, {Kuchner}, {Lane}, {Lillie},
  {Lyon}, {Lloyd}, {Lo}, {Lowrance}, {Macintosh}, {McCully}, {Marley},
  {Marois}, {Matthews}, {Mawet}, {Mazin}, {Mosier}, {Noecker}, {Pueyo},
  {Oppenheimer}, {Pedreiro}, {Postman}, {Roberge}, {Ridgeway}, {Schneider},
  {Schneider}, {Serabyn}, {Shaklan}, {Shao}, {Sivaramakrishman}, {Spergel},
  {Stapelfeldt}, {Tamura}, {Tenerelli}, {Tolls}, {Traub}, {Trauger},
  {Vanderbei}, \& {Wynn}}]{441}
{Levine}, M., {Soummer}, R., {Arenberg}, J., {et~al.} 2009, in Astronomy, Vol.
  2010, astro2010: The Astronomy and Astrophysics Decadal Survey, 37

\bibitem[{{Lindegren} \& {Dravins}(2003)}]{511}
{Lindegren}, L. \& {Dravins}, D. 2003, \aap, 401, 1185

\bibitem[{{Lovis} \& {Fischer}(2010)}]{490}
{Lovis}, C. \& {Fischer}, D. 2010, {Radial Velocity Techniques for Exoplanets}
  (University of Arizona Press), 27--53

\bibitem[{{Lyot}(1932)}]{483}
{Lyot}, B. 1932, \zap, 5, 73

\bibitem[{Mahalanobis(1936)}]{671}
Mahalanobis, P.~C. 1936, in Proceedings National Institute of Science, India,
  Vol.~2, 49--55

\bibitem[{{Mamajek} {et~al.}(2010){Mamajek}, {Kenworthy}, {Hinz}, \&
  {Meyer}}]{461}
{Mamajek}, E.~E., {Kenworthy}, M.~A., {Hinz}, P.~M., \& {Meyer}, M.~R. 2010,
  \aj, 139, 919

\bibitem[{{Mao} \& {Paczynski}(1991)}]{489}
{Mao}, S. \& {Paczynski}, B. 1991, \apjl, 374, L37

\bibitem[{{Marois} {et~al.}(2006){Marois}, {Lafreni{\`e}re}, {Doyon},
  {Macintosh}, \& {Nadeau}}]{480}
{Marois}, C., {Lafreni{\`e}re}, D., {Doyon}, R., {Macintosh}, B., \& {Nadeau},
  D. 2006, \apj, 641, 556

\bibitem[{{Mayor} \& {Queloz}(1995)}]{191}
{Mayor}, M. \& {Queloz}, D. 1995, \nat, 378, 355

\bibitem[{{McArthur} {et~al.}(2010){McArthur}, {Benedict}, {Barnes},
  {Martioli}, {Korzennik}, {Nelan}, \& {Butler}}]{401}
{McArthur}, B.~E., {Benedict}, G.~F., {Barnes}, R., {et~al.} 2010, \apj, 715,
  1203

\bibitem[{{Metropolis} {et~al.}(1953){Metropolis}, {Rosenbluth}, {Rosenbluth},
  {Teller}, \& {Teller}}]{550}
{Metropolis}, N., {Rosenbluth}, A.~W., {Rosenbluth}, M.~N., {Teller}, A.~H., \&
  {Teller}, E. 1953, \jcp, 21, 1087

\bibitem[{Moulton(1984)}]{673}
Moulton, F. 1984, An Introduction to Celestial Mechanics, Dover Books on
  Astronomy (Dover Publications)

\bibitem[{{Nascimbeni} {et~al.}(2011){Nascimbeni}, {Piotto}, {Bedin}, \&
  {Damasso}}]{446}
{Nascimbeni}, V., {Piotto}, G., {Bedin}, L.~R., \& {Damasso}, M. 2011, \aap,
  527, A85

\bibitem[{{OpenMP Architecture Review Board}(2008)}]{185}
{OpenMP Architecture Review Board}. 2008, {OpenMP}, http://openmp.org/

\bibitem[{{Perryman}(2000)}]{434}
{Perryman}, M.~A.~C. 2000, Reports on Progress in Physics, 63, 1209

\bibitem[{{Perryman} {et~al.}(1997){Perryman}, {Lindegren}, {Kovalevsky},
  {Hoeg}, {Bastian}, {Bernacca}, {Cr{\'e}z{\'e}}, {Donati}, {Grenon}, {van
  Leeuwen}, {van der Marel}, {Mignard}, {Murray}, {Le Poole}, {Schrijver},
  {Turon}, {Arenou}, {Froeschl{\'e}}, \& {Petersen}}]{241}
{Perryman}, M.~A.~C., {Lindegren}, L., {Kovalevsky}, J., {et~al.} 1997, \aap,
  323, L49

\bibitem[{{Pickering}(1890)}]{458}
{Pickering}, E.~C. 1890, The Observatory, 13, 80

\bibitem[{{Prevot}(1961)}]{290}
{Prevot}, L. 1961, Journal des Observateurs, 44, 83

\bibitem[{{Reffert}(2009)}]{491}
{Reffert}, S. 2009, \nar, 53, 329

\bibitem[{{Roberts}(1996)}]{150}
{Roberts}, G.~O. 1996, in Markov Chain Monte Carlo in Practice, 1st edn., ed.
  W.~R. {Gilks}, S.~{Richardson}, \& D.~J. {Spiegelhalter} (London: Chapman \&
  Hall), 45--57

\bibitem[{{Schneider}(2012)}]{192}
{Schneider}, J. 2012, {The Extrasolar Planets Encyclop{\ae}dia},
  http://exoplanet.eu/catalog.php

\bibitem[{{Seager}(2008)}]{444}
{Seager}, S. 2008, \ssr, 135, 345

\bibitem[{{Shao} {et~al.}(1988){Shao}, {Colavita}, {Hines}, {Staelin}, \&
  {Hutter}}]{296}
{Shao}, M., {Colavita}, M.~M., {Hines}, B.~E., {Staelin}, D.~H., \& {Hutter},
  D.~J. 1988, \aap, 193, 357

\bibitem[{Silverman(1986)}]{404}
Silverman, B. 1986, {Density estimation for statistics and data analysis},
  Monographs on statistics and applied probability (Chapman and Hall)

\bibitem[{Sivia(2006)}]{102}
Sivia, D.~S. 2006, Data Analysis---A Bayesian Tutorial, 2nd edn. (Oxford:
  Oxford University Press)

\bibitem[{Smith(1980)}]{600}
Smith, R.~L. 1980, in Bulletin of the TIMS/ORSA Joint National Meeting,
  Washington, DC, 101

\bibitem[{{Smith}(1987)}]{486}
{Smith}, W.~H. 1987, \pasp, 99, 1344

\bibitem[{{Sozzetti}(2005)}]{030}
{Sozzetti}, A. 2005, \pasp, 117, 1021

\bibitem[{{Thiele}(1883)}]{500}
{Thiele}, T.~N. 1883, Astronomische Nachrichten, 104, 245

\bibitem[{{Tolbert}(1964)}]{380}
{Tolbert}, C.~R. 1964, \apj, 139, 1105

\bibitem[{{van Leeuwen}(2007)}]{456}
{van Leeuwen}, F., ed. 2007, Astrophysics and Space Science Library, Vol. 350,
  {Hipparcos, the New Reduction of the Raw Data} (Springer Verlag)

\bibitem[{Vigan {et~al.}(2010)Vigan, Moutou, Langlois, Allard, Boccaletti,
  Carbillet, Mouillet, \& Smith}]{487}
Vigan, A., Moutou, C., Langlois, M., {et~al.} 2010, Monthly Notices of the
  Royal Astronomical Society, 407, 71

\bibitem[{{Vogt} {et~al.}(2005){Vogt}, {Butler}, {Marcy}, {Fischer}, {Henry},
  {Laughlin}, {Wright}, \& {Johnson}}]{011}
{Vogt}, S.~S., {Butler}, R.~P., {Marcy}, G.~W., {et~al.} 2005, \apj, 632, 638

\bibitem[{{Wolszczan} \& {Frail}(1992)}]{190}
{Wolszczan}, A. \& {Frail}, D.~A. 1992, \nat, 355, 145

\end{thebibliography}
	\bibliographystyle{aa}

\end{document}